\documentclass[11pt,english]{article}
\usepackage{graphicx}
\usepackage{graphics}
\usepackage{epsfig}
\usepackage{authblk}
\usepackage{subfigure}
\usepackage{wrapfig}
\usepackage[T1]{fontenc}
\usepackage[latin9]{inputenc}
\usepackage{units}
\usepackage{amsmath}
\usepackage{amssymb}
\usepackage{esint}
\usepackage{babel}
\usepackage{float}
\usepackage{empheq}
\usepackage{subfigure}
\usepackage{wrapfig}

\usepackage{xcolor}
\usepackage[normalem]{ulem}

\usepackage{setspace}

\title{ \small{{\it Published in Aerospace Journal, 2021}} \\ \vspace{6mm}
\LARGE{Numerical Investigation of a Rectangular Jet Exhausting over a Flat Plate with Periodic Surface Deformations at the Trailing Edge}}

\author[1]{Colby Horner \thanks{email: colbyhorner@hotmail.com}}
\author[1]{Adrian Sescu \thanks{email: sescu@ae.msstate.edu}}
\author[2]{Mohammed Afsar}
\author[3]{Eric Collins}

\affil[1]{Department of Aerospace Engineering, Mississippi State University, MS 39762, USA; colbyhorner@hotmail.com}
\affil[2]{Department of Mechanical \& Aerospace Engineering, Strathclyde University, Glasgow, UK}
\affil[3]{Center for Advanced Vehicular Systems, Mississippi State University, MS 39762, USA}

\begin{document}

\date{}

\maketitle

\begin{abstract}

Multiple competing factors are forcing aircraft designers to reconsider the underwing engine pod configuration typically seen on most modern commercial aircraft. One notable concern is increasing environmental regulations on noise emitted by aircraft. In an attempt to satisfy these constraints while maintaining or improving vehicle performance, engineers have been experimenting with some innovative aircraft designs which place the engines above the wings or embedded in the fuselage. In one configuration, a blended wing concept vehicle utilizes rectangular jet exhaust ports exiting from above the wing ahead of the trailing edge. While intuitively one would think that this design would reduce the noise levels transmitted to the ground due to the shielding provided by the wing, experimental studies have shown that this design can actually increase noise levels due to interactions of the jet exhaust with the aft wing surface and flat trailing edge. In this work, we take another look at this rectangular exhaust port configuration with some notional modifications to the geometry of the trailing edge to determine if the emitted noise levels due to jet interactions can be reduced with respect to a baseline configuration. We consider various horizontal and vertical offsets of the jet exit with respect to a flat plate standing in for the aft wing surface. We then introduce a series of sinusoidal deformations to the trailing edge of the plate of varying amplitude and wave number. Our results show that the emitted sound levels due to the jet--surface interactions can be significantly altered by the proposed geometry modifications. While sound levels remained fairly consistent over many configurations, there were some that showed both increased and decreased sound levels in specific directions. We present results here for the simulated configurations which showed the greatest decrease in overall sound levels with respect to the baseline. These results provide strong indications that such geometry modifications can potentially be tailored to optimize for further reductions in sound levels.

\end{abstract}


\section{Introduction}

Methods to reduce the noise generated by aircraft propulsion systems have long been a topic of research within the aerospace community. There have been many challenges faced by designers attempting to solve these types of problems due to the complexity of the flow phenomena as well as the costs involved with testing potential solutions. In some cases, noise reductions have only come at the expense of aircraft performance in other metrics (e.g., weight, cost).

Jet--surface interaction noise is usually associated with an increase in low frequency noise as a result of the interaction between a turbulent jet and a flat surface that is parallel to the jet axis. Jet noise is generally considered to be the result of two sources: scrubbing noise and trailing edge scattering noise (Brown and Wernet~\cite{Brown2}). Scrubbing is generated by pressure fluctuations in the turbulent boundary layer impinging on a flat surface. Trailing edge scattering, which is the dominant noise source at low frequency (Brown~\cite{Brown1}, Podboy~\cite{Podboy}), is produced when the hydrodynamic component of the upstream turbulence is scattered into acoustic waves after interacting with the trailing edge.
The total installation noise can propagate upstream or downstream of the trailing edge.

High-speed jet exhausts interacting with nearby surfaces can generate flow distortions that significantly increase the overall noise emitted. In one such example, an aircraft engine placed on top of the wing in an attempt to shield the jet noise that would propagate to the ground while also improving aircraft performance actually resulted in an overall increase in sound levels. Another example of note is the tremendous noise generated by the jet exhausts of military aircraft interacting with the various surfaces of an aircraft carrier deck during take-off and landing operations.

Considerable research has been conducted on axisymmetric jets (Bridges~\cite{Bridges2}, Brown and Bridges~\cite{Brown4}), but, since the 1980s, a number of studies have shifted their focus to non-circular jet engine exit designs as potential passive flow control devices (Gutmark and Grinstein~\cite{Gutmark}). Geometric corrugations applied at the nozzle exit plane (e.g., tabbed studied by Gao et al.~\cite{Gao}, cruciform by El Hassan et al.~\cite{El_Hassan}, and chevrons by Violato et al.~\cite{Violato}) have been observed to break down the large-scale structures present in the jet exhaust flow. This subsequently reduces the mixing between the jet and ambient-fluid. It was noted that these effects were most pronounced for larger wetted perimeters of the jet exit (Shakouchi and Iryama~\cite{Shakouchi}) and an increased number of corners/sides (Gutmark and Grinstein~\cite{Gutmark}). {The effect of external expansion ramps on supersonic flow emitted from high-aspect-ratio rectangular nozzles was analyzed by Malla et al.~\cite{Malla} to determine if the ramps could be used to reduce noise.} {Research conducted by Mancinelli et al.~\cite{Mancinelli} and Proenca et al.~\cite{Proenca} show further results regarding the interaction of rectangular jets with flat plates.} Other aircraft propulsion concepts have utilized acoustically treated engine parts (inlet ducts, exhaust ducts, and inner walls), but this type of treatment tends to have a negative overall impact due to additional cost and weight.

While static chevron configurations at the jet exit have been shown to provide a significant reduction in noise without noticeable loss in thrust, some researchers have been looking into potential acoustic benefits that could be gained by exciting the chevrons with piezoelectric actuators. Mechanical perturbations of the flow by the actuated chevrons are believed to increase the production of small-scale disturbances while diminishing the large-scale turbulent structures thought to be responsible for the dominant portion of jet mixing noise. Mohan et al.~\cite{Mohan} introduced piezoelectric actuators to chevrons in an attempt to modify the growth rate of the mixing layer. Butler and Calkins~\cite{Butler} examined four different types of nozzles that exist: round, faceted, faceted with static chevrons, and faceted with active chevrons. Their results demonstrated a 2 to 4 dB reduction in noise with the static chevrons, and an additional 2 dB reduction with the actuated chevron configurations.

Zaman~\cite{Zaman3} studied the integration of microjets near the engine exhaust port and found that a clear noise reduction was observed as the microjet pressure increased. The results showed that turbulent mixing noise reduction, as monitored by the overall sound pressure level at a shallow angle, correlated with the ratio of the microjet to the primary jet driving pressures (normalized by the ratio of the corresponding diameters). {Semlitsch et~al.~\cite{Semlitsch} used implicit large eddy simulation to analyze the relationship between screech tone frequency and fluidic injection pressure of the microjets.}

Rego et al.~\cite{Rego} utilized a Lattice--Boltzmann Method (LBM) to carry out numerical simulations of a flat plate placed in an irrotational hydrodynamic field of a jet with a round nozzle. Three cases were investigated with input flow characteristics determined from NASA wind tunnel experiments (see Brown~\cite{Brown3}). Heated and cooled jets were analyzed at different Mach numbers, and the far-field noise was computed using the Ffowcs-Williams and Hawkings equation. Far-field spectral results showed a large noise increase for low to mid frequencies due to upstream hydrodynamic waves in the jet that propagate along the plate, 'scrubbing' over the surface, until it reaches the trailing edge of the plate. (This is referred to as the 'gust solution' in Afsar et al.~\cite{Afsar}). As these waves interact with the trailing edge and the downstream turbulent mixing layer, they are scattered in the far-field as noise. A noticeable change in the far-field noise was observed when the trailing edge location was changed relative to the fixed nozzle. As the length of the plate was increased, the trailing edge was positioned in a region dominated by large-scale structures and a greater degree of low-frequency amplification was observed. When the plate was moved closer to the jet in the radial direction, the sound levels at lower frequencies increased. 

Behrouzi and McGuirk~\cite{Behrouzi} conducted experiments at the Loughborough University high-pressure nozzle test facility on a rectangular nozzle exhausting over a rectangular plate acting as an aft-deck. (A configuration similar to the one used in the current study.) The focus of the experiment was to study the effect of the rectangular plate on flow field development and to determine if any acoustic benefits could be obtained from the shielding effect that the aft-deck might produce. Schlieren imaging, pneumatic Pitot probes, and nonintrusive Laser Doppler Anemometry measurements were used to capture detailed information on the flow structures. The data revealed that the aft-deck created an asymmetry in the entrainment characteristics of the shear layers, inducing a net transverse pressure force on the plume by changing the inviscid shock cell structure. The presence of the aft-deck altered the net pressure force exerted on the flow which led to a transverse deflection of the jet. The aft-deck configuration produced a dramatic effect on the plume development, extending the potential core length slightly and reducing turbulence levels in the plume near field.


Bridges~\cite{Bridges} further investigated this configuration by studying whether beveled edges on the rectangular nozzle would make a difference in the sound generated when compared to an aft-deck being added to the propulsion configuration (see also Zaman~\cite{Zaman} and Zaman et al.~\cite{Zaman2}). Nozzles were fabricated with different aspect ratios of 2:1, 4:1, and 8:1. For each aspect ratio, three nozzle designs were tried: one basic nozzle with no extension and two beveled variants with bevel lengths of 1.3 and 2.7 inches. A rectangular nozzle with an aft-deck surface was also considered in which the rectangular nozzles had plates fitted to them such that when the plate surface was even with the inner lip of the nozzle, the surface and nozzle were effectively one piece. Five aft-deck lengths were tested: 1.3, 2.7, 4, 8, and 12 inches long. When the beveled nozzle was analyzed, it was found that the noise levels decreased in both azimuthal planes as the aspect ratio was increased, with much more dramatic reductions in the nozzle's major axis plane. These reductions in noise were lost when one lip of the nozzle was extended to make an aft deck. Low and high frequencies were examined with the aft deck extensions. For low frequencies, the sound levels were roughly the same on both sides of the plate. However, there were noticeable differences in the sound levels for higher frequencies on each side of the plate. This is due to the effect of the noise being shielded and reflected. The shielding effect seemed to increase as the length of the aft deck increased, but this was not the case for reflection. In addition, it is important to note that, for the largest plate lengths, the low-frequency amplification above the baseline jet noise was not dependent on the aspect ratio.

Seiner and Manning~\cite{Seiner} considered the interaction between a supersonic jet and a flat plate from a rectangular nozzle. They showed that the distance between the nozzle exit and flat surface was an important parameter which can impact the screech noise. Ibrahim et~al.~\cite{Ibrahim} studied the effect of turbulence characteristics of jet flows on the radiated jet noise. {Berland et al.~\cite{Berland} used compressible large eddy simulations to investigate the generation of screech tones from an under-expanded jet with a three-dimensional planar geometry.}


In the present work, we aim to extend these results by examining the effect of sinusoidal surface deformations added to the trailing edge of the aft-deck plate. To this end, a suite of large eddy simulations have been performed, each targeting different geometrical configurations and flow conditions. Sound levels in the farfield were then computed to determine whether these deformations could potentially reduce the total noise emitted by the installed configuration. In these simulations, we consider a high aspect ratio rectangular jet. To simplify the nozzle geometry, we used a spanwise slice of width four times the height of the nozzle with periodic boundary conditions applied in the spanwise direction. Flow disturbances were imposed at the inflow boundaries to introduce three-dimensional flow structures into the jet flow. A high-order accurate solver was used to discretize the unsteady, compressible, conservative form of the filtered Navier--Stokes equations. The mean flow was determined using Reynolds Averaged Navier--Stokes (RANS) equations via a {SST} $k-\omega$ turbulence model {(Menter~\cite{Menter})}. The farfield noise radiation was evaluated using Ffowcs-Williams and a Hawkings acoustic analogy method.


We provide further details of the numerical formulation in Section~\ref{numerical-formulation}, and present our results in Section~\ref{results}.


\section{Problem Formulation and Numerical Algorithms}\label{numerical-formulation}
\subsection{Scalings}%

\begin{figure}[htp]
 \begin{center}
    \begin{tabular}{c}
    \includegraphics[width=0.65\textwidth]{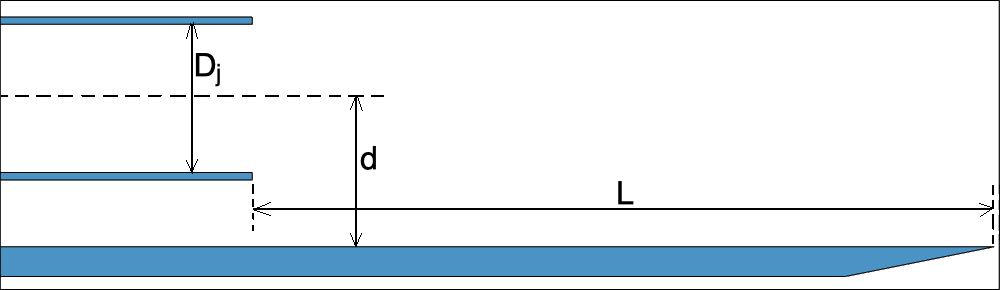} \\
    (\textbf{a}) \\
    \includegraphics[width=0.45\textwidth]{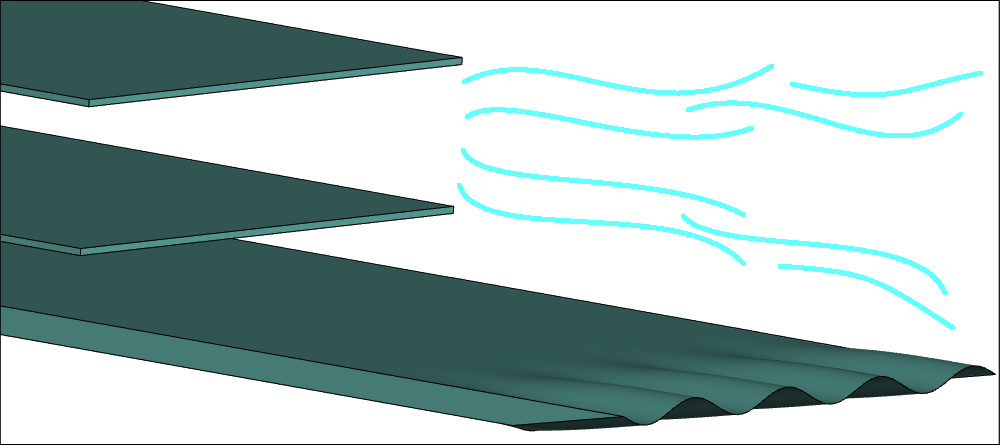} 
    \includegraphics[width=0.45\textwidth]{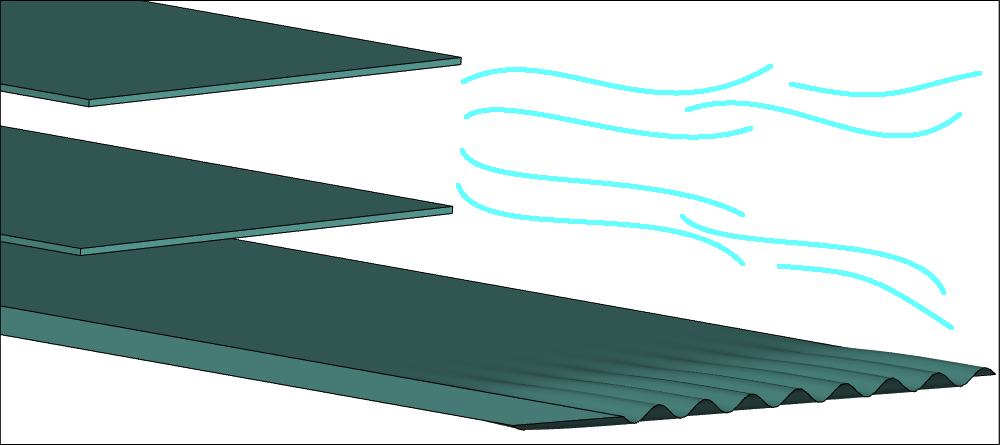} \\
    (\textbf{b}) \hspace{63mm}  (\textbf{c})
     \end{tabular}
 \end{center}
    \caption{(\textbf{a}) The geometry of the nozzle and the plate; (\textbf{b},\textbf{c}) two examples of trailing edge deformations of different wavenumbers.}
    \label{f1}
\end{figure}

The governing equations considered here are expressed in terms of a generalized curvilinear coordinate system, subject to the following transformations:
\begin{eqnarray}\label{curvilinear-coords}
\xi = \xi \left(x,y,z \right), \hspace{0.2in}
\eta = \eta \left(x,y,z \right), \hspace{0.2in}
\zeta = \zeta \left(x,y,z \right), \nonumber
\end{eqnarray}
where $\xi$, $\eta$, and $\zeta$ are the body-aligned curvilinear coordinates corresponding to the streamwise, wall-normal, and spanwise directions, respectively, and $x$, $y$, and $z$ are the (non-dimensional) global Cartesian coordinates of physical space. All dimensional spatial coordinates ($x^*,y^*,z^*$) are normalized by the reference length, $D_j$, the height of the nozzle (See Figure~\ref{f1})
\begin{eqnarray}\label{scaled-coords}
(x,y,z) = \frac{(x^*,y^*,z^*)}{D_j},
\end{eqnarray}

The fluid velocity values are scaled by the {jet} velocity, $V_j$.
\begin{eqnarray}\label{scaled-velocity}
(u,v,w) = \frac{(u^*,v^*,w^*)}{V_j},
\end{eqnarray}

Pressure values are scaled by the dynamic pressure, $\rho_{\infty} V_j^{2}$, and temperature values by the freestream temperature, $T_{\infty}$. Reynolds number, Mach number, and Prandtl number are defined as:
\begin{eqnarray}\label{nondimensional-quantities}
Re = \frac{\rho_{\infty} V_j D_{j}}{\mu_{\infty}}, \hspace{5mm}
Ma = \frac{V_j}{a_{\infty}^*}, \hspace{5mm}
Pr = \frac{\mu_{\infty} C_p}{k_{\infty}}
\end{eqnarray}
where $\mu_{\infty}$, $a_{\infty}$, and $k_{\infty}$ are the freestream dynamic viscosity, speed of sound, and thermal conductivity, respectively. $C_p$ is the specific heat at constant pressure. For all simulations, these values are initialized for air as an ideal gas.

\subsection{Governing Equations}%

The filtered Navier--Stokes equations can be expressed in conservative 
 form as:
\begin{eqnarray}\label{Navier--Stokes}
\mathbf{Q}_t
+ \mathbf{F}_{\xi}
+ \mathbf{G}_{\eta}
+ \mathbf{H}_{\zeta}
= \mathbf{S}.
\end{eqnarray}

Here, the subscripts denote partial derivatives with respect to time and the curvilinear coordinates. $\mathbf{Q}$ is the vector of conserved quantities:
\begin{equation}
\mathbf{Q} = \frac{1}{J} \left \lbrace \rho,\; \rho \vec{u},\; E \right \rbrace^{T},
\end{equation}
$\rho = \rho^*/\rho_\infty$ is the non-dimensional density of the fluid, $\vec{u} = (u, v, w)$ is the non-dimensional velocity vector in physical space, and $E$ is the total energy. The flux vectors, $\mathbf{F}$, $\mathbf{G}$ and $\mathbf{H}$, are given by:
\begin{eqnarray}
\mathbf{F} = \frac{1}{J} \left\{ 
\begin{array}{c}
\rho U \\
\rho \vec{u} U + \xi_{x_i} (p + \tau_{i1}) \\
E U + p \tilde{U} +  \xi_{x_i} \Theta_i
\end{array}
\right\},  \nonumber \\
\mathbf{G} = \frac{1}{J} \left\{ 
\begin{array}{c}
\rho V \\
\rho \vec{u} V + \eta_{x_i} (p + \tau_{i2}) \\
E V + p \tilde{V}+  \eta_{x_i} \Theta_i
\end{array}
\right\},  \nonumber  \\ 
\mathbf{H} = \frac{1}{J} \left\{ 
\begin{array}{c}
\rho W \\
\rho \vec{u} W + \zeta_{x_i} (p + \tau_{i3}) \\
E W + p \tilde{W}+  \zeta_{x_i} \Theta_i
\end{array}
\right\},  \nonumber
\end{eqnarray}

Here, the repeated indices indicate Einstein summation convention over $i\in{1,2,3}$. $U, V, W$ are the contravariant velocity components,
\begin{eqnarray}\label{con}
U = \xi_x u + \xi_y v + \xi_z w, \\
V = \eta_x u + \eta_y v + \eta_z w, \\
W = \zeta_x u + \zeta_y v + \zeta_z w,
\end{eqnarray}
$\tau$ is the shear stress tensor,
\begin{equation}
\tau_{ij} = \frac{\mu}{Re} \left[
\left(
\frac{\partial \xi_k}{\partial x_j}  \frac{\partial u_i}{\partial \xi_k}  +
\frac{\partial \xi_k}{\partial x_i}  \frac{\partial u_j}{\partial \xi_k}
\right)
- \frac{2}{3} \delta_{ij} \frac{\partial \xi_l}{\partial x_k}  \frac{\partial u_k}{\partial \xi_l}
\right],
\end{equation}
and $\Theta$ is the heat flux,
\begin{equation}
\Theta_{i} = 
 u_j \tau_{ij} + \frac{\mu}{(\gamma-1)M_{\infty}^2 Re Pr}
\frac{\partial \xi_l}{\partial x_i}  \frac{\partial T}{\partial \xi_l}
\end{equation}

Again, repeated indices (that are not present on the left-hand side) indicate summation. The Jacobian of the curvilinear transformation from the physical space to computational space is denoted by $J$, and $\mathbf{S}$ is a vector of prescribed source terms. Pressure, temperature, and density are related by the ideal gas equation of state:
\begin{equation}\label{eos}
p = \frac{\rho T}{\gamma M_{\infty}^2}
\end{equation}

The dynamic viscosity and thermal conductivity $k$ are related to temperature using the Sutherland's equations in dimensionless form:
\begin{eqnarray}
\mu = T^{3/2} \frac{1 + C_1/T_{\infty}}{T+C_1/T_{\infty}}; \hspace{4mm}
k = T^{3/2} \frac{1 + C_2/T_{\infty}}{T+C_2/T_{\infty}},
\end{eqnarray}

For air at sea level, $C_1 = 110.4$ K, and $C_2 = 194$ K.

There are no explicit subgrid scale turbulence terms in Equation~(\ref{Navier--Stokes}) {(see Grinstein et al.~\cite{Grinstein})}. Instead, the compressible Navier--Stokes equations are solved within the framework of an implicit large eddy simulation, where numerical filtering is applied to account for the missing sub-grid scale energy. The numerical solver uses high-order finite difference approximations for the spatial derivatives and explicit time marching. The time integration is performed using a second order Adams--Bashforth method~\cite{Butcher}:
\begin{equation}\label{32}
\mathbf{Q}^{n+1} = \mathbf{Q}^n + k
\left[
\sum_{\nu=0}^{K}\beta_\nu L(\mathbf{Q}^{n-\nu})
\right]
\end{equation}

Here, the constants $\beta_\nu$ are chosen to give either the maximum order of accuracy~\cite{Butcher} or the lowest dispersion and dissipation. $L(\mathbf{Q})$ is the residual.

The spatial derivatives are discretized using the dispersion-relation-preserving schemes of Tam and Webb~\cite{Tam} or a high-resolution 9-point dispersion-relation-preserving optimized scheme of Bogey and Bailly~\cite{Bogey}. To damp out the unwanted high wavenumber waves from the solution, high-order spatial filters, as developed by Kennedy and Carpenter~\cite{Kennedy}, are used. No slip boundary conditions for velocity and adiabatic conditions for temperature are imposed at the solid surfaces. Sponge layers are imposed near the far-field boundaries in regions that are outside the flow domain of interest. The sponge layers combined with grid stretching act to damp-out unwanted waves returning from the farfield boundaries. For the LES simulations, the {flow in proximity to the} wall is modeled using the Werner--Wengler model (Werner and Wengle~\cite{Werner}), {so some small turbulent flow structures may not captured adequately}. At the inflow boundary, {a constant mean flow in the core region is combined with a hyperbolic tangent function in the shear region, and} disturbances in the form of a superposition of Fourier modes with random amplitudes, frequencies, and wavenumbers. {These disturbances introduce three-dimensional flow variations into the jet~flow}.


The mean flows used to initialize the LES are obtained from RANS simulations, where a classical {SST} $k-\omega$ turbulence model (Menter~\cite{Menter}) is applied to account for missing fluctuations. {This model was found to perform well for jet applications (see Mihaescu et~al.~\cite{Mihaescu})}

In the present work, high-order, central-difference schemes are used to achieve increased resolution of the propagating disturbances. A shock-capturing technique suitable for simulations involving central differences in space is required to avoid unwanted oscillations that may propagate from discontinuities arising in supersonic flows. Shock capturing techniques are employed based on the general explicit filtering framework, a straightforward approach which introduces sufficient numerical viscosity in the area of the discontinuities, and negligible artificial viscosity in the rest of the domain. The technique of Bogey et al.~\cite{Bogey2} introduces selective filtering at each grid vertex to minimize numerical oscillations, and shock-capturing in the areas where discontinuities are present. This method has been proven to work efficiently for high-order accurate, nonlinear computations.



\subsection{Ffowcs-Williams Hawkings Acoustic Analogy Method}%

The Ffowcs-Williams and Hawkings (FW-H) equation \cite{Brentner,Ffowcs} is an inhomogeneous wave equation that can be derived by manipulating the continuity equation and the Navier--Stokes equations for a compressible fluid. The FW--H equation can be written as:
\begin{eqnarray}\label{a0}
&& \frac{1}{a_0^2}\frac{\partial^2 p'}{\partial t^2} - \nabla^2 p' = \frac{\partial^2}{\partial x_i \partial x_j}\{ T_{ij} H(f) \} \nonumber \\
&-&\frac{\partial}{\partial x_i} \{ \left[ P_{ij}n_j + \rho u_i (u_n-v_n) \right] \delta(f) \}  \nonumber  \\
&+& \frac{\partial}{\partial t} \{ \left[ \rho_0 v_n + \rho (u_n-v_n) \right] \delta(f) \}
\end{eqnarray}
where $u_i$ is the fluid velocity component in the  $x_i$ direction, $u_n$ is the fluid velocity component normal to the surface at $f=0$, $v_i$ is the surface velocity components in the $x_i$ direction, $v_n$ is the surface velocity component normal to the surface, $\delta(f)$ is Dirac delta function, $H(f)$is Heaviside function, and $p'=p-p_0$ is the magnitude of the sound pressure disturbance at the far field.

The acoustic pressure in the farfield at a location, $\vec{\mathbf{x}}$, can be determined from:
\begin{equation}\label{a1}
p'(\vec{\mathbf{x}},t)    = p'_T(\vec{\mathbf{x}},t) + p'_L(\vec{\mathbf{x}},t) + p'_Q(\vec{\mathbf{x}},t)
\end{equation}
where
\begin{equation}\label{a2}
p'_T(\vec{\mathbf{x}},t) = \frac{1}{4\pi} \frac{\partial}{\partial t} \int_{S} \left[ \frac{\rho_0 U_n}{r} \right]_{ret} dS
\end{equation}
\begin{equation}\label{a3}
p'_L(\vec{\mathbf{x}},t) = \frac{1}{4\pi a_0} \frac{\partial}{\partial t} \int_{S} \left[ \frac{L_r}{r} \right]_{ret} dS  
                       + \frac{1}{4\pi} \int_{S} \left[ \frac{L_r}{r^2} \right]_{ret} dS
\end{equation}

Given a point, $\vec{\mathbf{y}}$, in the near-field acoustic source on surface $S$, $r = |\vec{r}| = |\vec{\mathbf{x}} - \vec{\mathbf{y}}|$ is the distance from the surface source location to the farfield position $\vec{\mathbf{x}}$; $U_n$ is the scalar product of
$ 
\vec{U} = \rho \vec{u}/\rho_0
$ 
and the unit normal to the surface $S$ at $\vec{\mathbf{y}}$; $L_r$ is the scalar product of
$ 
L_i = P_{ij} n_i + \rho u_i u_n
$ 
and the unit vector in the direction from $\vec{\mathbf{y}}$ to $\vec{\mathbf{x}}$;
$ 
P_{ij} = -\tau_{ij} + p \delta_{ij}
$ 
; $\tau_{ij}$ is the viscous stress; $\rho$ is the instantaneous density; and $\rho_0$ and $a_0$ are the ambient density and sound speed, respectively.

In Equation (\ref{a1}), $p'_Q(\vec{\mathbf{x}},t)$ is the contribution from the quadrupole, but it is neglected here. The integrals in (\ref{a2}) and (\ref{a3}) are calculated at the emission time $\tau = t - r/a$, where $a$ is the speed of sound and $t$ is the reception time. The flow variables at $\tau$ must be interpolated from the flow data at multiple time iterations, $t$, due to its dependency on $\vec{\mathbf{y}}$, which varies over the surface $S$ surrounding the jet.

For the numerical implementation of the FW-H method, a conical surface that includes the entire jet is considered. The primitive variables and viscous stresses are interpolated from the flow domain to the surface at every $b$ time steps (the value of $b$ depends on the desired time resolution) and written out to a large number of files for post-processing. Given a reception time $t$, the emission time $\tau$ is calculated for every small surface element $\Delta S$ on the conical surface. Within a specific surface element $\Delta S$ on the surface, an interpolation is necessary to obtain the flow data at the emission time $\tau$, which is needed to calculate the integrals in (\ref{a2}) and (\ref{a3}). Next, we determine the time derivatives of $U_n$ and $L_r$ and evaluate the integrals in Equations (\ref{a2}) and (\ref{a3}). The acoustic pressure then is calculated from Equation (\ref{a1}).

\section{Results and Discussion}\label{results}

Results are reported and discussed in this section for a high aspect ratio rectangular jet exhausting over a flat surface located underneath the jet flow. Because the aspect ratio of the nozzle exit is high, we simplify the problem by considering only a portion from the nozzle along the spanwise direction. This simplification is based on the assumption that the flow is statistically two-dimensional (an assumption also considered by~\cite{Zaman2}). Thus, the spanwise length of the flow domain is set to be four times the height of the nozzle, and periodic boundary conditions are imposed in the lateral direction.

The layout of the configuration is illustrated schematically in Figure~\ref{f1}a. The flat plate is oriented parallel to the jet axis and located beneath the jet exit. The vertical separation of the plate from the jet axis ($d$) and the distance from the jet exit to the trailing edge of the plate ($L$) were both varied in this study. These distances are specified below and in Table~\ref{t1} in terms of multiples of the nozzle height ($D_j$). The vertical separation considered were 0.65, 0.75, and 1 times the nozzle height, while the horizontal distances to the trailing edge considered were 4, 5, and 6 times the nozzle height. A 3D view of the geometry of the nozzle and the plate with deformed trailing edge are shown in Figure~\ref{f1}. The deformations in the spanwise direction are prescribed as a sine function while the amplitude is modulated by a hyperbolic tangent function in the streamwise direction. Two combinations of wavenumbers and amplitudes are shown in Figure~\ref{f1}b,c.

The Reynolds number, based on the jet velocity and the height of the nozzle, is 500,000, and the acoustic Mach number is $0.8$. The amplitude of velocity disturbances imposed at the nozzle exit are restricted to less than 3\% of the jet velocity.

 \begin{table}[htp]
 \caption{Position of the plate with respect to the nozzle (see also Figure~\ref{f1}).}
  \label{t1}
 \begin{center}
\begin{tabular}{ccc}  \hline
      \textbf{Case} &  \textbf{Vertical Stand-Off Dist. (d)} & \textbf{Streamwise Dist. from the Nozzle Exit (L) }   \\  \hline
       1 &  0.75 $D_j$ &  4 $D_j$   \\
       2 &  0.75 $D_j$ &  5 $D_j$  \\ 
       3 &  0.75 $D_j$ &  6 $D_j$    \\
       4 &  1.0 $D_j$ &  5 $D_j$     \\ 
       5 &  0.65   $D_j$ &  5 $D_j$     \\   \hline
         \end{tabular}
 \end{center}
 \end{table}

\subsection{RANS Results}\label{s3}

\begin{figure}[htp]
 \begin{center}
 \begin{tabular}{c}
    \includegraphics[width=0.5\textwidth]{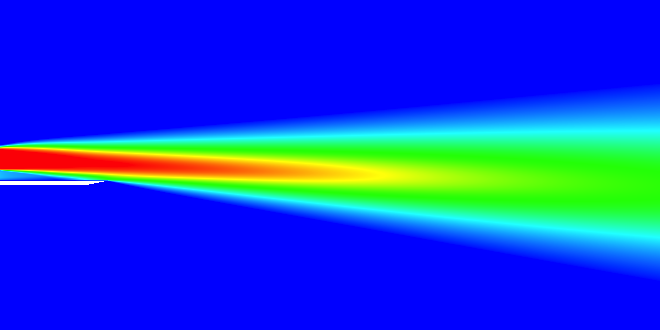}
    \includegraphics[width=0.5\textwidth]{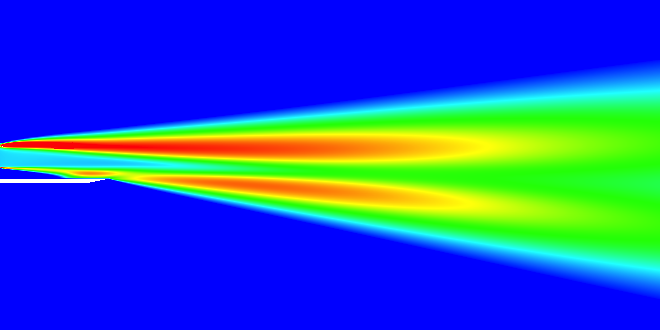} \\
   (\textbf{a}) \hspace{83mm}  (\textbf{b}) \\
    \includegraphics[width=0.5\textwidth]{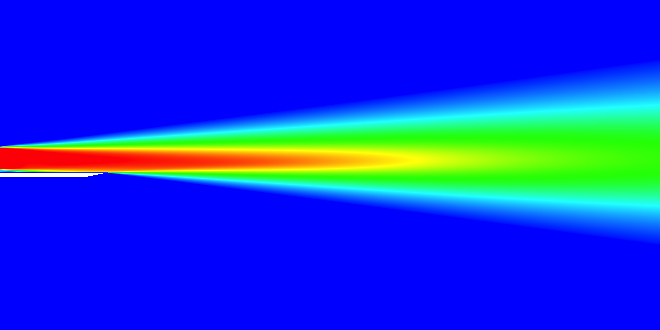}
    \includegraphics[width=0.5\textwidth]{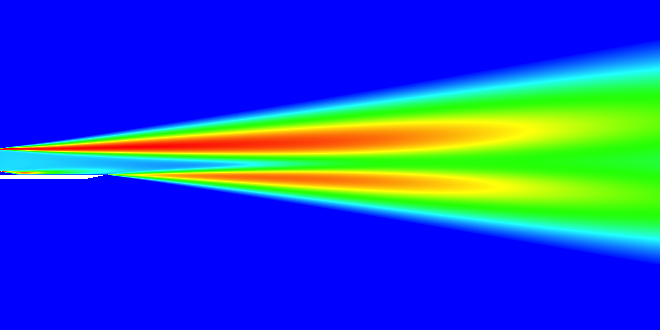} \\
    (\textbf{c}) \hspace{83mm}  (\textbf{d}) \\
    \end{tabular}
 \end{center}
    \caption{Mean velocity magnitude (\textbf{left}) and turbulent kinetic energy (\textbf{right}) distributions from RANS simulations of two cases. Top (\textbf{a},\textbf{b}): $d = 1.0 D_j$, $L = 5 D_j$; Bottom (\textbf{c},\textbf{d}): $d = 0.65 D_j$, $L = 5 D_j$.}
    \label{f2}
\end{figure}

\begin{figure}[htp]
 \begin{center}
	\subfigure[]{\includegraphics[width=0.45\textwidth]{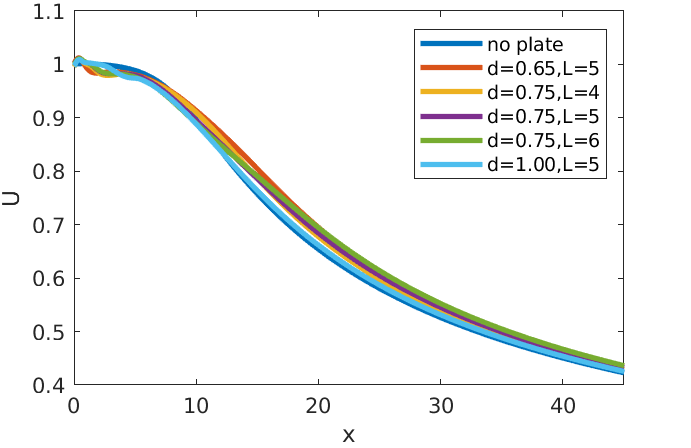}}
        \subfigure[]{\includegraphics[width=0.46\textwidth]{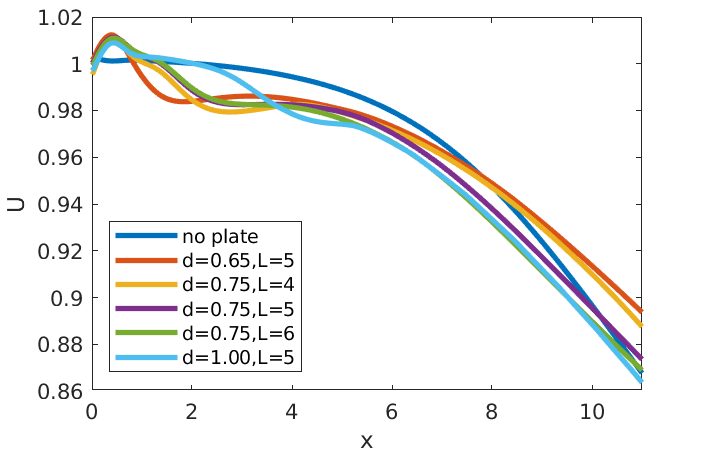}}
	\caption{Mean streamwise velocity along the jet centerline from RANS: (\textbf{a}) $x$ from $0$ to $45$; (\textbf{b}) details in the core region with $x$ from $0$ to $12$.}
	\label{f3}
 \end{center}
\end{figure}

\begin{figure}[htp]
 \begin{center}
	\subfigure[]{\includegraphics[width=0.45\textwidth]{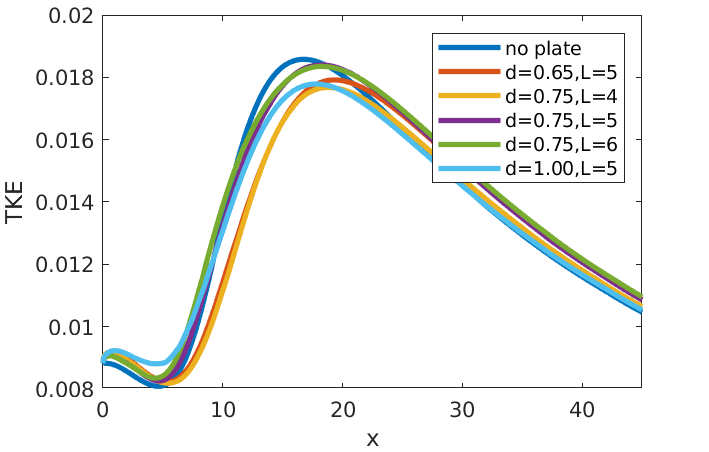}}
	\subfigure[]{\includegraphics[width=0.45\textwidth]{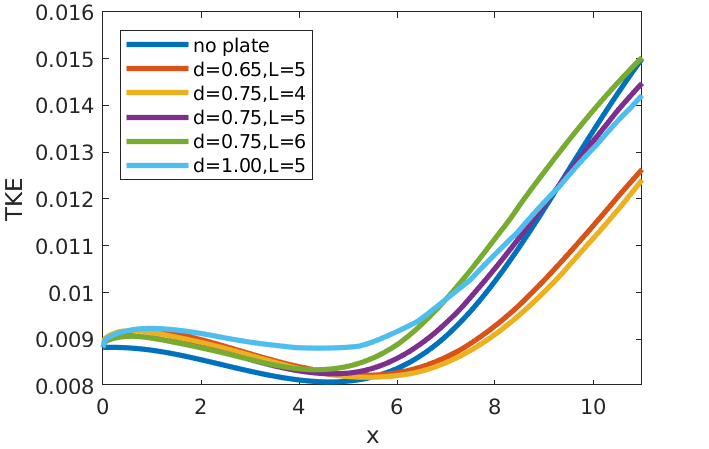}}
	\caption{Turbulent kinetic energy along the jet centerline from RANS: (\textbf{a}) $x$ from $0$ to $45$; (\textbf{b}) details in the core region with $x$ from $0$ to $12$.}
	\label{f4}
 \end{center}
\end{figure}

\begin{figure}[htp]
 \begin{center}
\begin{tabular}{cc}
	\subfigure[]{\includegraphics[width=0.45\textwidth]{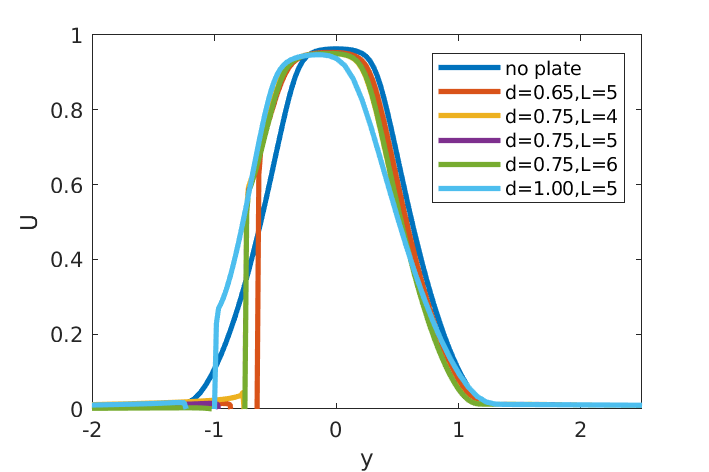}} &
	\subfigure[]{\includegraphics[width=0.45\textwidth]{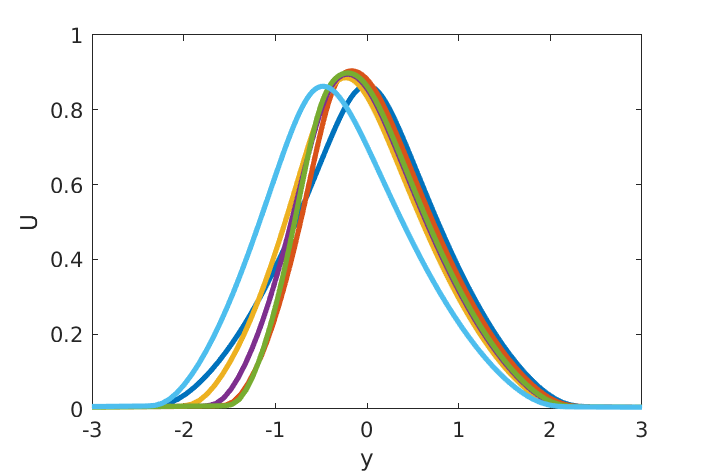}}\\
	\subfigure[]{\includegraphics[width=0.45\textwidth]{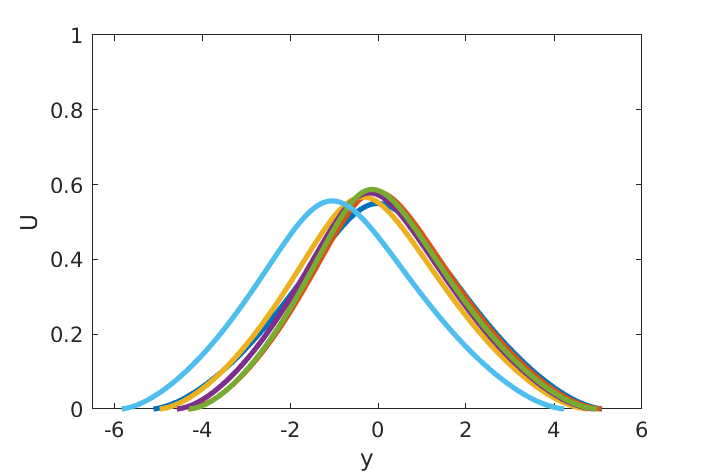}} &
	\subfigure[]{\includegraphics[width=0.45\textwidth]{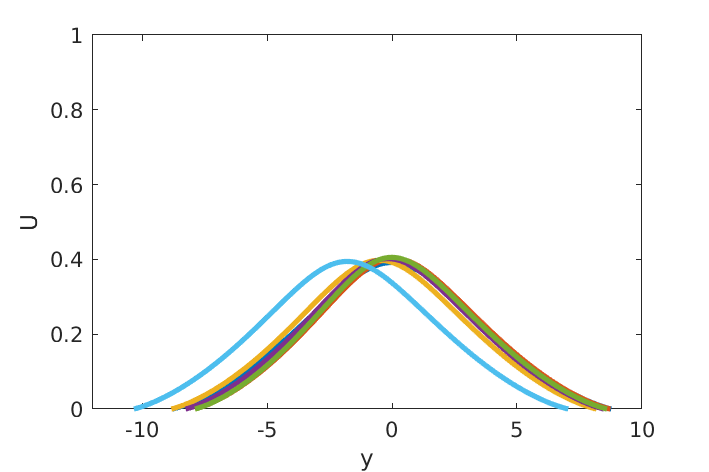} }
	\end{tabular}
	\caption{Streamwise velocity 
	profiles along vertical direction, for different configurations: (\textbf{a})~$x~=~4 D_j$; (\textbf{b}) $x = 10 D_j$; (\textbf{c}) $x = 25 D_j$; (\textbf{d}) $x = 45 D_j$.}
	\label{f8}
 \end{center}
\end{figure}

\begin{figure}[htp]
 \begin{center}
	\subfigure[]{ \includegraphics[width=0.45\textwidth]{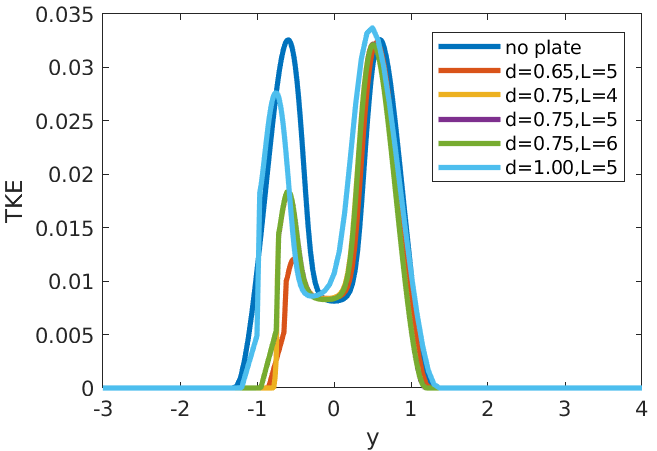}} 
	\subfigure[]{\includegraphics[width=0.49\textwidth]{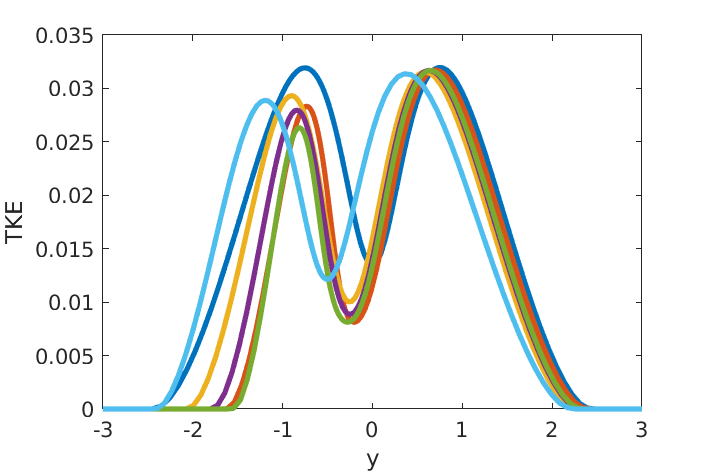} }
	\subfigure[]{\includegraphics[width=0.45\textwidth]{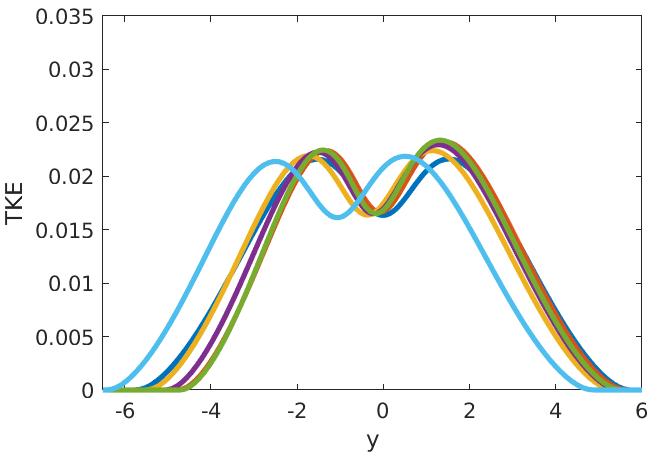} }
	\subfigure[]{\includegraphics[width=0.45\textwidth]{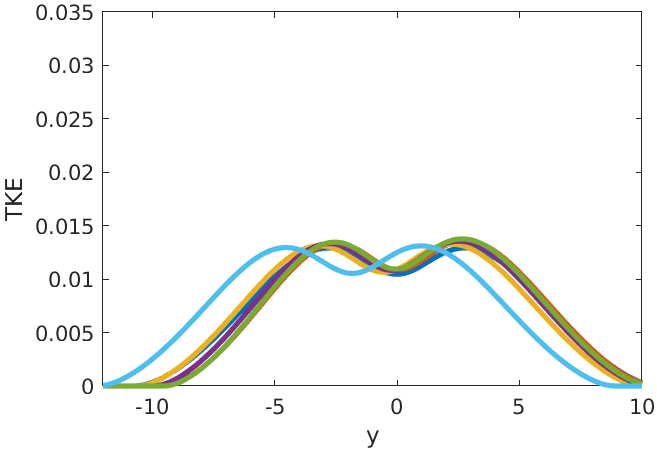} }
	\caption{Turbulent kinetic 
	energy profiles along vertical direction, for different configurations: (\textbf{a})~$x = 4 D_j$; (\textbf{b}) $x = 10 D_j$; (\textbf{c}) $x = 25 D_j$; (\textbf{d}) $x = 45 D_j$.}
	\label{f9}
 \end{center}
\end{figure}

The mean flow and turbulent kinetic energy (TKE) used to initialize the LES are obtained from separate two-dimensional RANS simulations using a
$\kappa-\omega$ turbulence model. The RANS results shown in Figures \ref{f2}--\ref{f9} are reported first to demonstrate the effect of the plate location on the jet mean flow and TKE. Plots of mean velocity magnitude (left) {and turbulent kinetic energy (right)} are shown in Figure~\ref{f2} for {two selected cases, corresponding to the largest and the smallest offset distance of the plate from the jet axis.} These color plots indicate that there is a visible effect from the plate on {both} the mean flow {and the turbulent kinetic energy}. For the first configuration ($d = 1.0 D_j$, $L = 5 D_j$), which corresponds to the largest distance $d$ from the center of the nozzle, the jet is significantly deviated downward by the plate {(resembling the 'Coanda' effect)}. This deviation is also present in other cases, but they do not appear to be {as significant as it is for this case}. The smallest deviation {from the jet axis occurs} for the configuration ($d = 0.65 D_j$, $L = 5 D_j$), which corresponds to the smallest offset distance, $d$.

{From the plots of TKE shown on the right side of Figure~\ref{f2},} there appears to be an attenuation of the TKE in the bottom shear layer of the jet as a result of the interaction with the plate boundary layer. This is potentially due to the viscous dissipation in the boundary layer that is developing on the surface of the plate. Further downstream of the plate trailing edge, the magnitude of the TKE appears to recover somewhat as the two shear layers show similar intensities. This is further revealed by the profiles of TKE across the jet shown in Figure~\ref{f9}.

\vspace{-6pt}
\vspace{-6pt}
\newpage
Center-line mean velocity distributions are plotted in Figure~\ref{f3} for several tested configurations along with the results obtained from the baseline free jet (no plate). The presence of the plate seems to slow down the flow inside the potential core as seen in the zoomed-in plot shown in Figure~\ref{f3}b. While all curves seem to follow the same trend in the downstream region, there is a slight acceleration for the cases that include the plate. This does seem to depend on the distance of the jet axis to the plate, $d$; the smaller this distance, the higher the acceleration. TKE distributions along the jet center line that are plotted in Figure~\ref{f4} reveal a decrease of TKE in the downstream region for the cases that involve the plate. This decrease is more significant for the configurations with the smallest $d$ and smallest $L$ (yellow and red curves in Figure~\ref{f4}).

In Figures~\ref{f8} and \ref{f9}, we compare cross-flow vertical profiles of mean velocity magnitude and turbulent kinetic energy among the five configurations, at four axial locations, $x = 4 D_j$;, $x = 10 D_j$, $x = 25 D_j$, and $x = 45 D_j$; the results from the free jet case are also included. They all show that the largest deviation of the jet from the original position occurs for $d = 1.0 D_j$ and $L = 5 D_j$, corresponding to the largest distance of the plate from the jet (this was also observed in the contour plots of the mean velocity and TKE). Figure \ref{f9} shows that the reduction of the TKE in the lower shear layer is more significant for the configuration with the smallest distance from the trailing edge. Figures \ref{f8}d and \ref{f9}d indicate that the jet flow becomes more symmetric further downstream for all normalized profiles, except the case corresponding to $d = 1.0 D_j$, $L = 5 D_j$.

\subsection{LES Results}\label{s4}


A suite of LES runs were carried out corresponding to the geometrical configurations given in Table \ref{t1}, and the trailing edge deformations characterized by the function
\begin{eqnarray}
g(x,z) = 0.5 A ((1+\tanh(\sigma_x (x+x_0))) * \sin(2 \pi (wz) z/L_z)
\end{eqnarray}
 where $x_0=-1.2$ is the streamwise location on the plate where the deformation begins to gradually increase from zero to the maximum amplitude at the trailing edge location; $\sigma_x=2$ is a parameter controlling the 'thickness' of the tangent hyperbolic function; and $L_z$ is the spanwise width of the plate. The amplitude, $A$, and wavenumber, $k$, are varied as given in Table~\ref{t2}. (See also Figure~\ref{f1} for a visual representation of the trailing edge deformations.) As mentioned previously, the spanwise length of the flow domain is four times the height of the nozzle, and periodic boundary conditions are imposed in the lateral direction with the assumption that the aspect ratio of the rectangular nozzle is very high. This lateral dimension of $4D_j$ was selected to ensure that all relevant turbulent flow structures are captured accurately in the spanwise direction. The mesh consists of approximately $10$~millions grid points, clustered mostly in proximity to the jet region. Grid stretching is used in the farfield along with an imposed sponge layer condition at the outflow boundary to ensure that the fluctuations are gradually dissipated while they leave the flow domain.

\begin{figure}[htp]
 \begin{center}
         \includegraphics[width=0.7\textwidth]{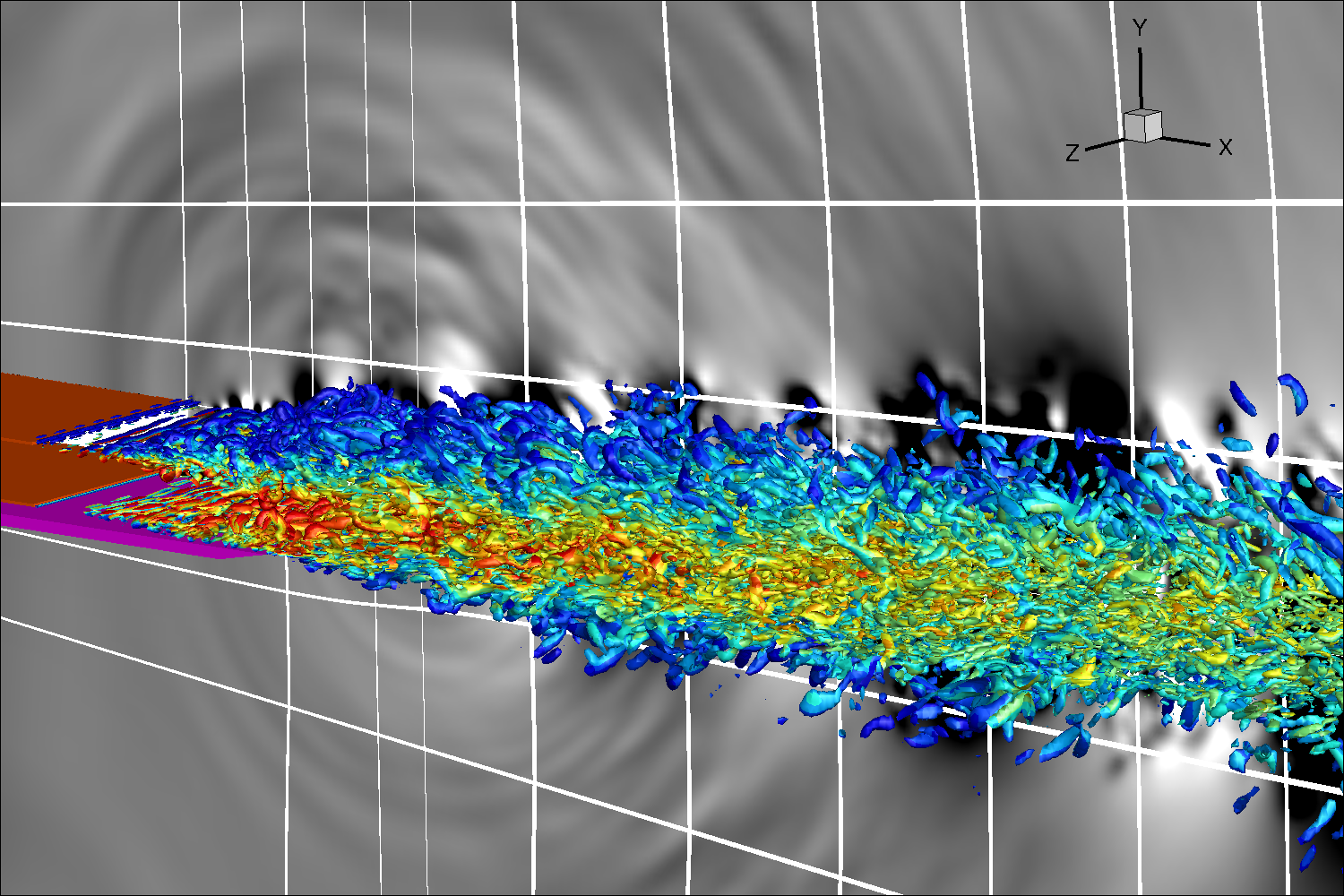}
	\caption{Iso-surfaces of Q-criterion colored by the streamwise velocity and pressure contours in~gray.}
	\label{f10}
 \end{center}
\end{figure}

 Figure \ref{f10} shows iso-surfaces of Q-criterion colored by the streamwise velocity component. ($Q = 1/2[|\mathbf{\Omega}|^2 - |\mathbf{S}|^2]$, where $\mathbf{S} = 1/2[\nabla \mathbf{v}+(\nabla \mathbf{v})^T]$ is the rate-of-strain tensor, and $\mathbf{\Omega} = 1/2[\nabla \mathbf{v}-(\nabla \mathbf{v})^T]$ is the vorticity tensor.) Acoustic waves radiating from the jet are plotted in gray contours. (The white lines are caused by gaps between blocks from the MPI decomposition). The attenuation of acoustic waves propagating from the jet underneath the plate can be observed. This is a result of the shielding provided by the plate causing the waves to not be as intense in this region. 

 \begin{table}[htp]
  \caption{Amplitudes and wavenumbers of the trailing edge.}
  \label{t2}
 \begin{center}
  \begin{tabular}{ccc}  \hline
      \textbf{Treatment} & \textbf{Amplitude, {\it A}} &  \textbf{Wavenumber, {\it wz}}    \\   \hline
       1 &  0.04 $D_j$ &  10    \\
       2 &  0.06 $D_j$ &  8   \\ 
       3 &  0.08 $D_j$ &  6     \\ 
       4 &  0.12 $D_j$ &  5      \\ 
       5 &  0.14 $D_j$ &  4      \\
       6 &  0.16 $D_j$ &  3      \\   \hline
  \end{tabular}
 \end{center}
 
 \end{table} 
\vspace{-5pt}

\begin{figure}[htp]
 \begin{center}
\begin{tabular}{ccc}
	\subfigure[]{\includegraphics[width=0.3\textwidth]{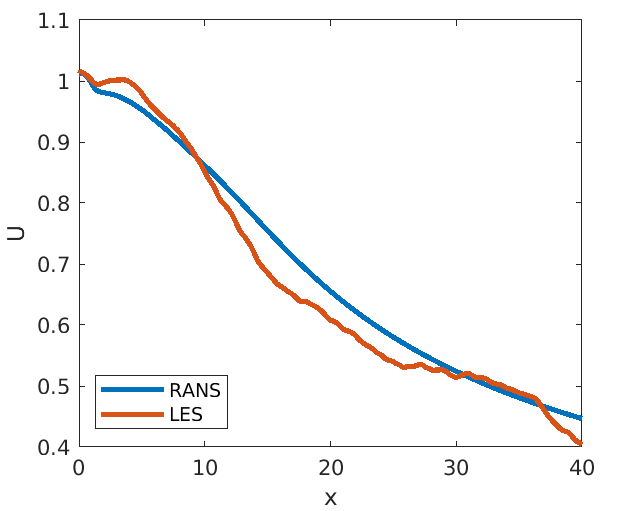}} &
	\subfigure[]{\includegraphics[width=0.3\textwidth]{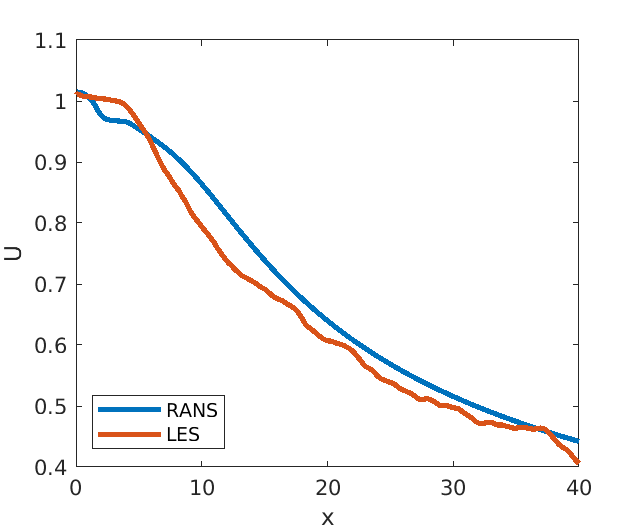}}&
	\subfigure[]{\includegraphics[width=0.3\textwidth]{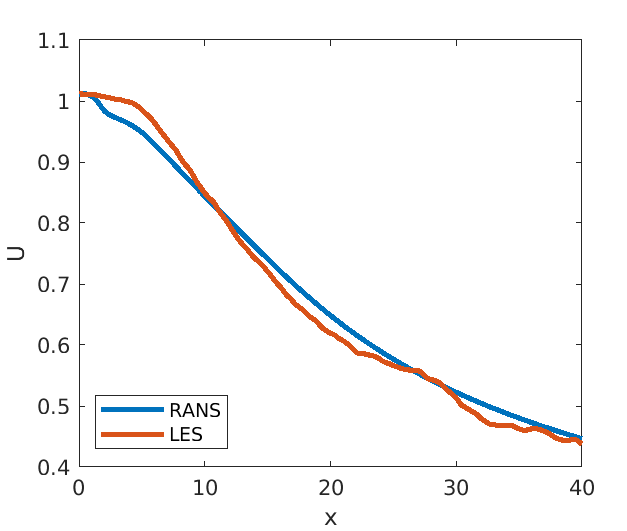}}\\
	\subfigure[]{\includegraphics[width=0.3\textwidth]{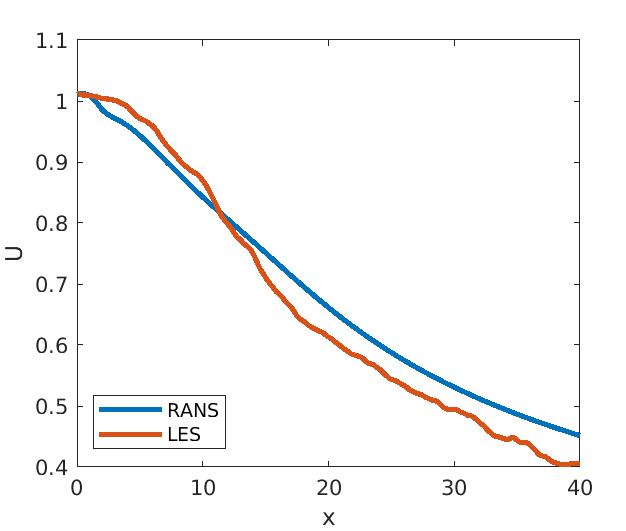}}&
	\subfigure[]{\includegraphics[width=0.3\textwidth]{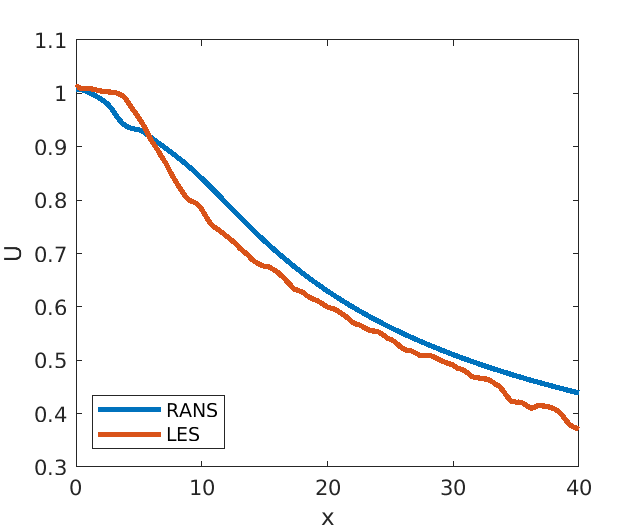}} &
	\subfigure[]{\includegraphics[width=0.3\textwidth]{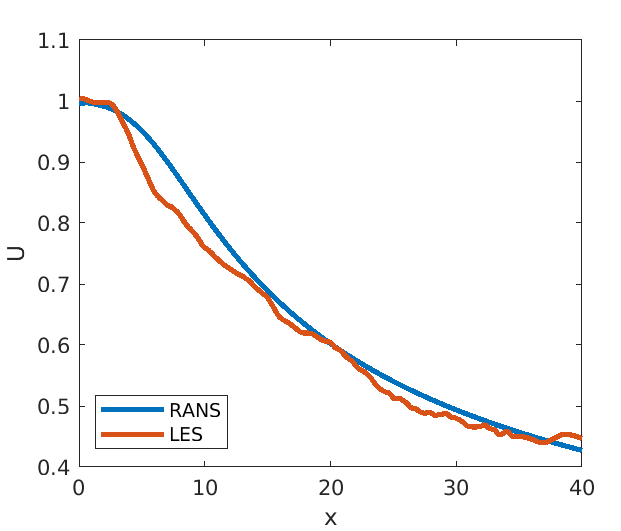}}
\end{tabular}	
	\caption{Mean streamwise velocity along the jet centerline: (\textbf{a})  $d = 0.65 D_j$, $L = 5 D_j$; (\textbf{b}) $d = 0.75 D_j$, $L = 4 D_j$; (\textbf{c}) $d = 0.75 D_j$, $L = 5 D_j$; (\textbf{d}) $d = 0.75 D_j$, $L = 6 D_j$; (\textbf{e}) $d = 1.0 D_j$, $L = 5 D_j$; (\textbf{f})  no plate.}
	\label{f11}
 \end{center}
\end{figure}

The mean centerline velocity distributions are compared against RANS results in Figure~\ref{f11}. While there are some slight differences between the two in the potential core region, overall the agreement is good, showing similar decay in the downstream flow. (The LES profiles show some fluctuations from insufficient time spans used for the time averaging).  

\begin{figure}[htp]
 \begin{center}
         \includegraphics[width=0.7\textwidth]{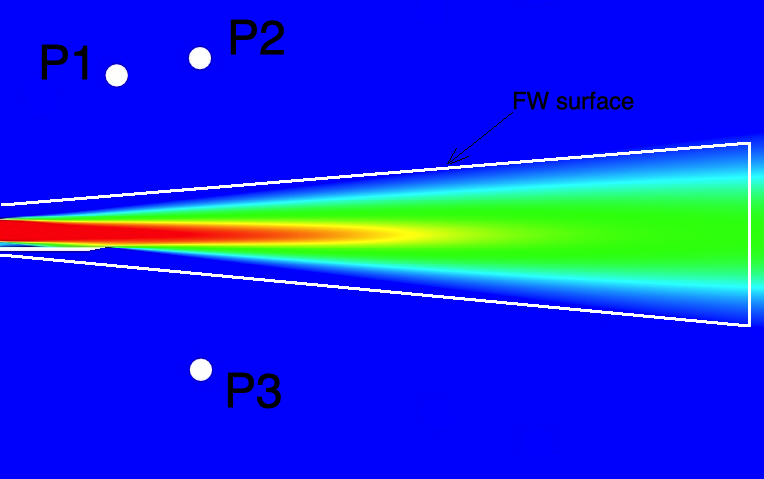}
	\caption{Locations of the three Probes, and the Ffowcs-Williams integration surface.}
	\label{f12}
 \end{center}
\end{figure}

Acoustic spectra have been calculated at three probe locations relatively close to the jet (in the nearfield). These probes are located at large angles with respect to the jet axis since it was observed that jet surface interaction noise dominates at large polar angles (see~\cite{Goldstein} for a study of AR8 nozzle and~\cite{Afsar} where various rectangular nozzle jets were considered). {The Ffowcs-Williams Hawkins surface is shown in Figure~\ref{f12}. The vertical side in the downstream was placed at 30 nozzle diameters from the nozzle exit plane. (Varying this distance did not affect the results significantly.)} Two probes are located above the jet while the third is located underneath the jet (see Figure~\ref{f12}). {A fourth probe was placed in the far field at 100 equivalent diameters and $50$ deg angle with respect to the jet axis to compare the numerical results with experimental results collected by Bridges et al.~\cite{Bridges3} (see also Bridges~\cite{Bridges4}).} Results were analyzed by using the sound pressure level and overall sound pressure level plots. The Strouhal number, $St_e$, was computed by the formula: 
\begin{equation}\label{StrouhalNumber}
St_e = \frac{f D_j}{V_j}
\end{equation}
where f is the frequency, $D_j$ is nozzle height, and $V_j$ is the jet velocity.

\begin{figure}[htp]
 \begin{center}
	\includegraphics[width=0.45\textwidth]{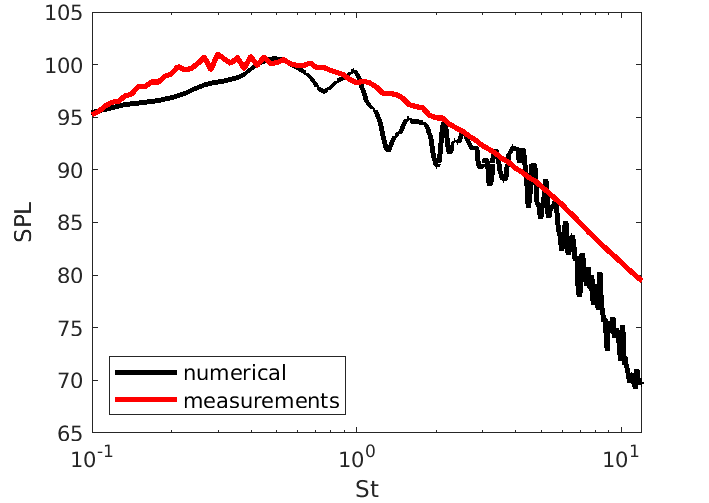}
	\caption{Comparison between 
	numerical results and measurements (Bridges et al.~\cite{Bridges3}).}
	\label{f_exp}
 \end{center}
\end{figure}

{Figure \ref{f_exp} shows the comparison between our acoustic spectrum and the spectrum obtained from measurements (Bridges et al.~\cite{Bridges3}). Bridges et al.~\cite{Bridges3} conducted an extensive set of measurements to acquire acoustic data on single-flow convergent rectangular nozzles of aspect ratios 2:1, 4:1, and 8:1, at Mach numbers 0.7 and 0.9. Our comparison is for the largest aspect ratio 8:1 and Mach number 0.9. The agreement in Figure~\ref{f_exp} is fairly good in the low frequency range of the spectrum. For the high frequency range the spectrum curve obtained from numerical simulations is slightly lower, which is most likely due to the mesh resolution of the LES not being sufficient to fully resolve the smallest flow scales that are responsible for the highest frequency noise.}

 \begin{table}[htp]
\caption{Probe locations.}
  \label{t3}
 \begin{center}
 \begin{tabular}{ccc}  \hline
      \textbf{Probe} & \textbf{X-Location} & \textbf{Y-Location}    \\  \hline
       1 &  0 &  5   \\ 
       2 &  1 &  5.3     \\ 
       3 &  1  &  $-$3      \\  \hline
  \end{tabular}
 \end{center}

 \end{table} 
 
\begin{figure}[htp]
 \begin{center}
 \begin{tabular}{cc}
	\subfigure[]{\includegraphics[width=0.45\textwidth]{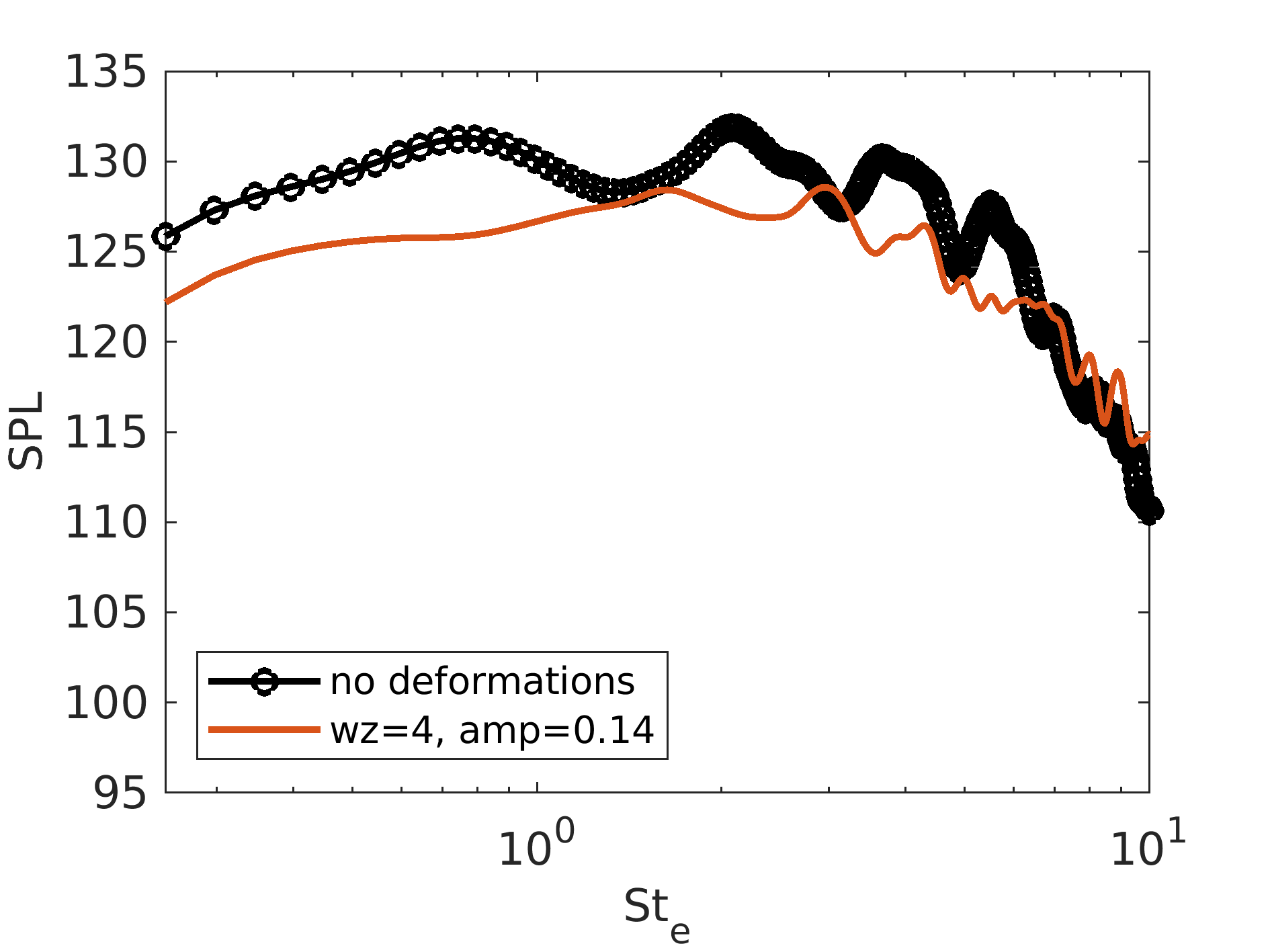}}&
	\subfigure[]{\includegraphics[width=0.45\textwidth]{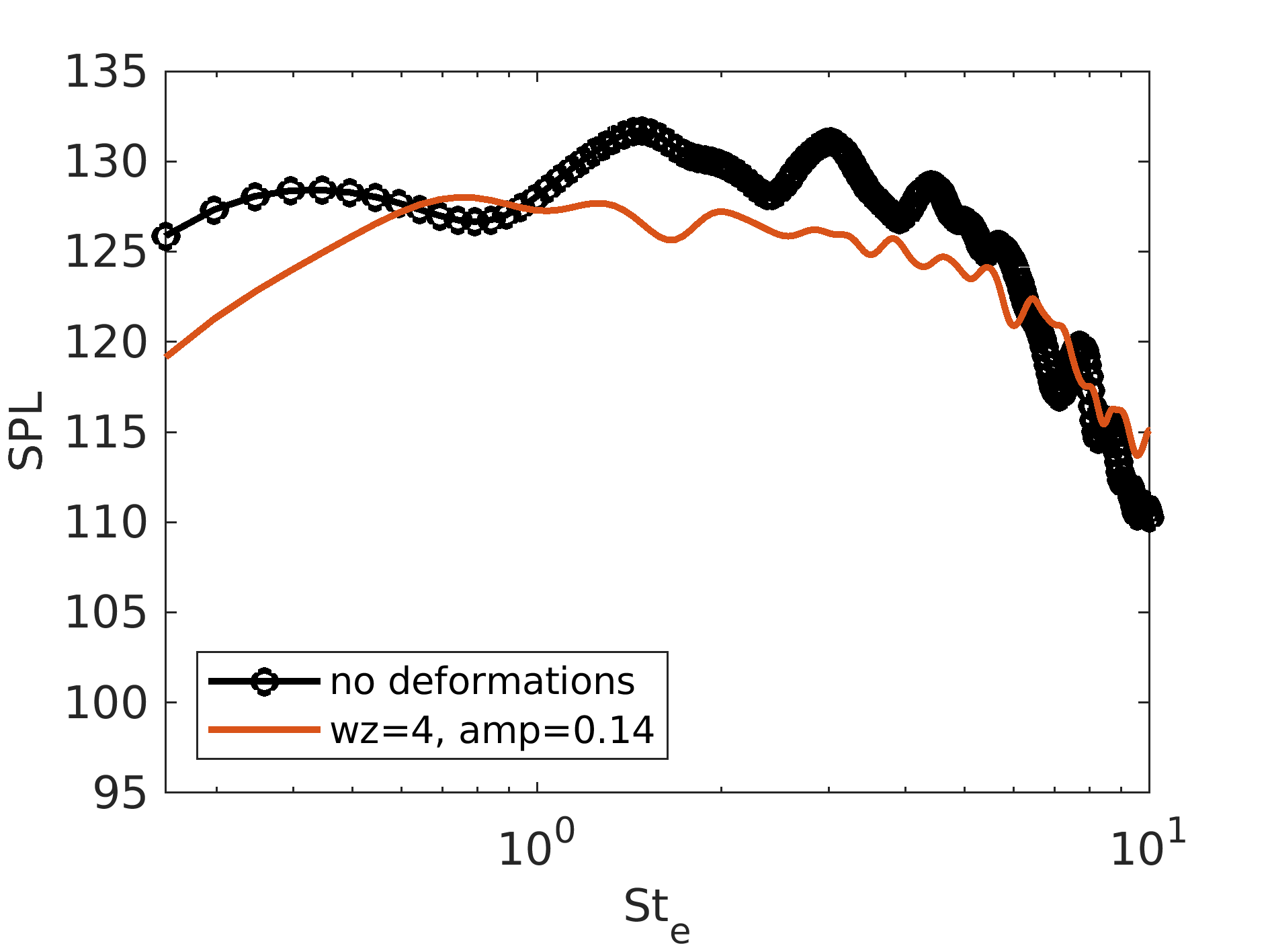}}\\
	\multicolumn{2}{c}{\subfigure[]{\includegraphics[width=0.45\textwidth]{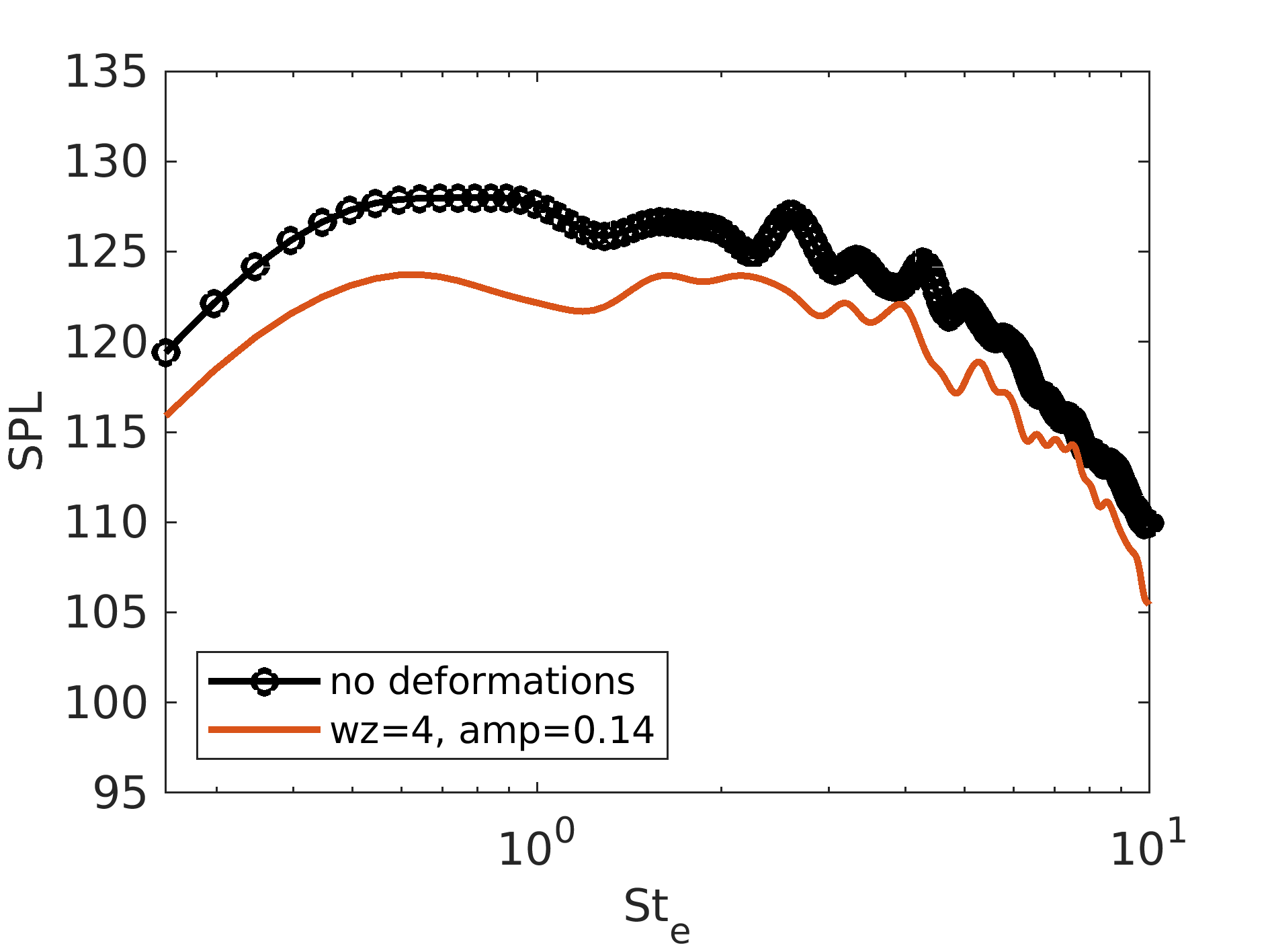}}}
	 \end{tabular}
	\caption{Typical acoustic spectra that show a reduction: (\textbf{a}) Probe 1; (\textbf{b}) Probe 2; (\textbf{c}) Probe 3.}
	\label{f15}
 \end{center}
\end{figure}

In general, the effect of trailing edge deformation on the radiated noise was small. We noticed both an increase and a decrease in noise, depending on the combination of amplitudes and wavenumbers characterizing the deformations. A visible reduction in noise was achieved for treatment 5 given in Table \ref{t2} ($w_z=4$, $A=0.14$). The acoustic spectra at the three probe points are plotted for this case in Figure \ref{f15}a--c.

\begin{figure}[htp]
 \begin{center}
\begin{tabular}{cc}
	\subfigure[]{\includegraphics[width=0.45\textwidth]{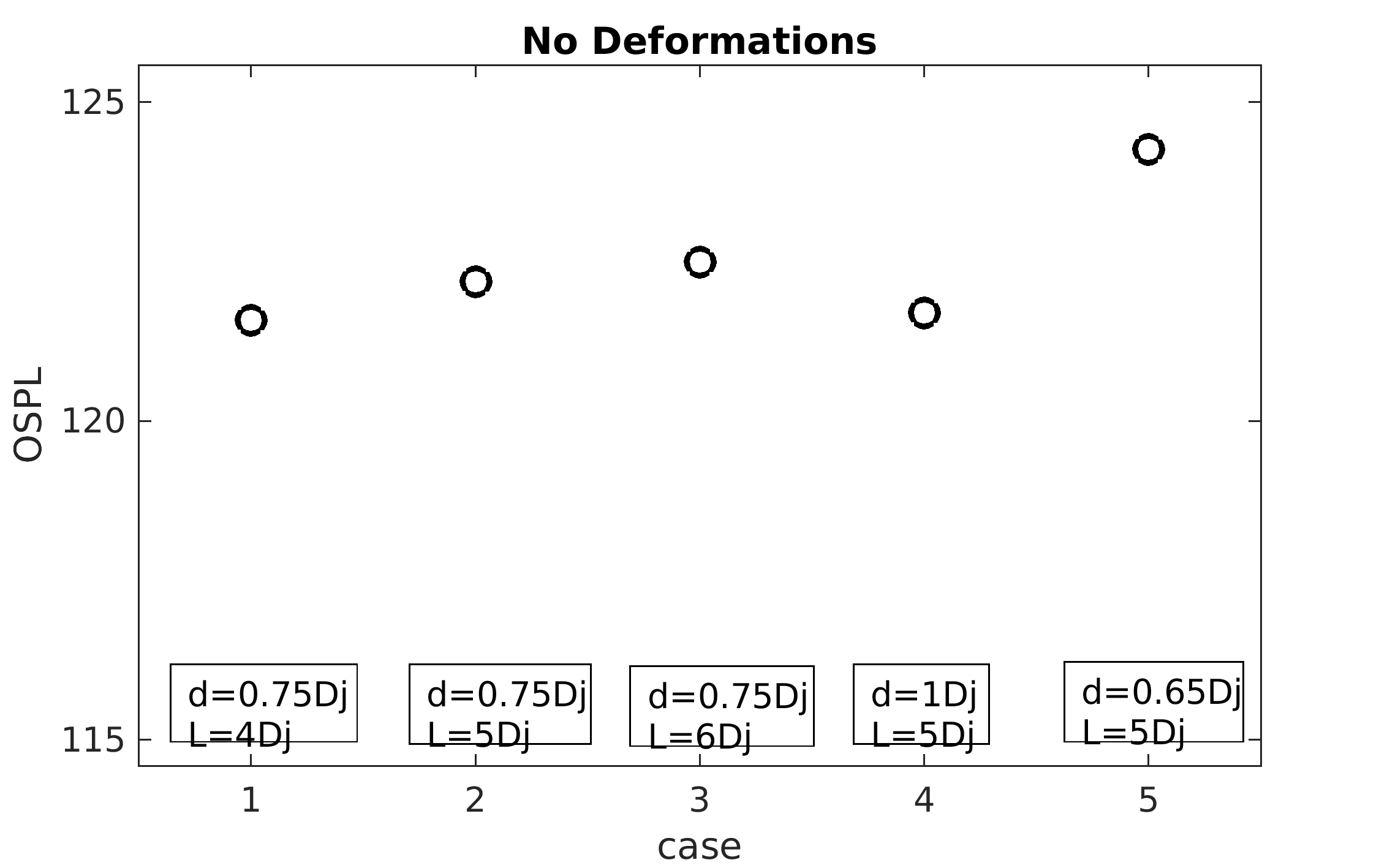} }&
	\subfigure[]{\includegraphics[width=0.45\textwidth]{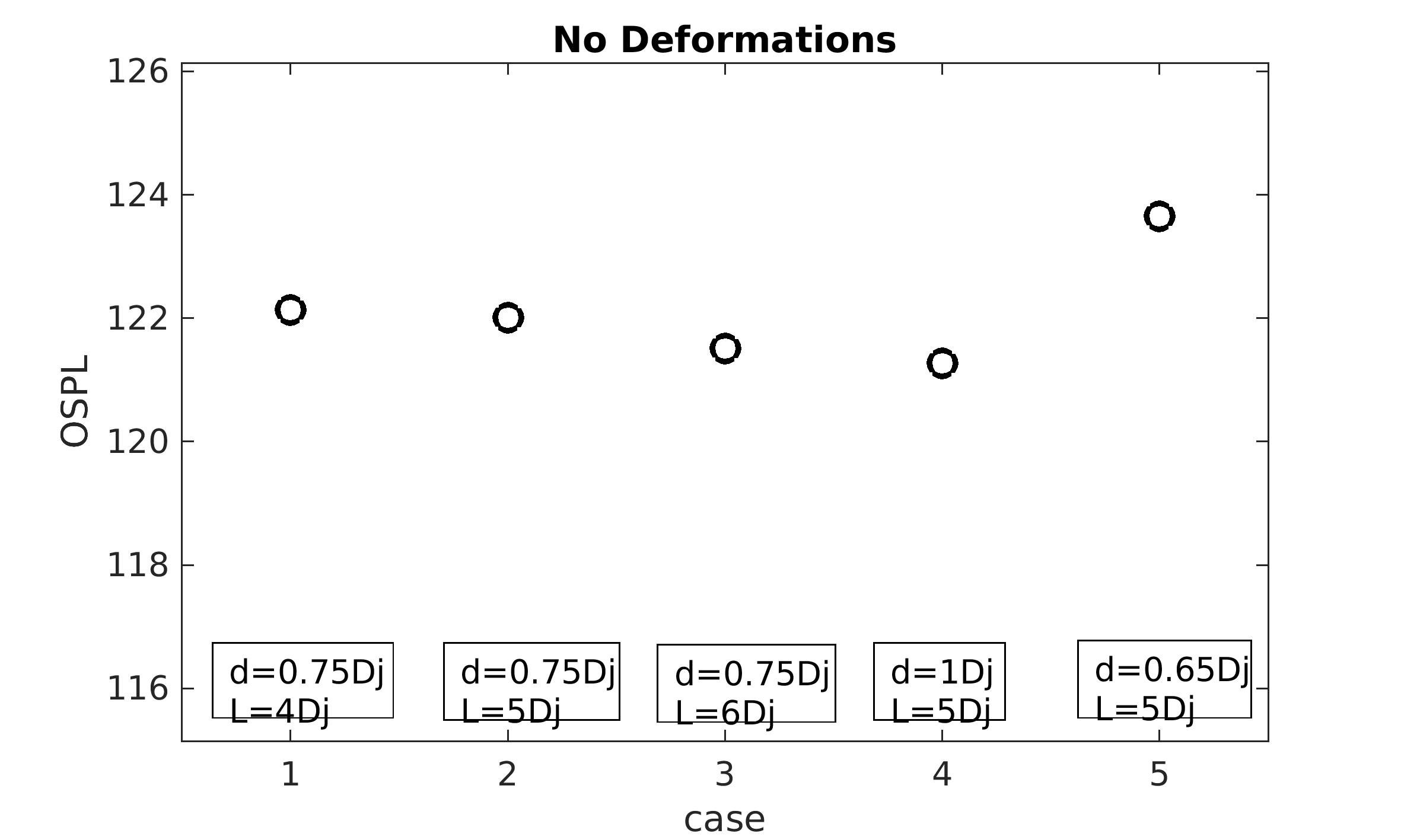}}\\
	\multicolumn{2}{c}{\subfigure[]{\includegraphics[width=0.45\textwidth]{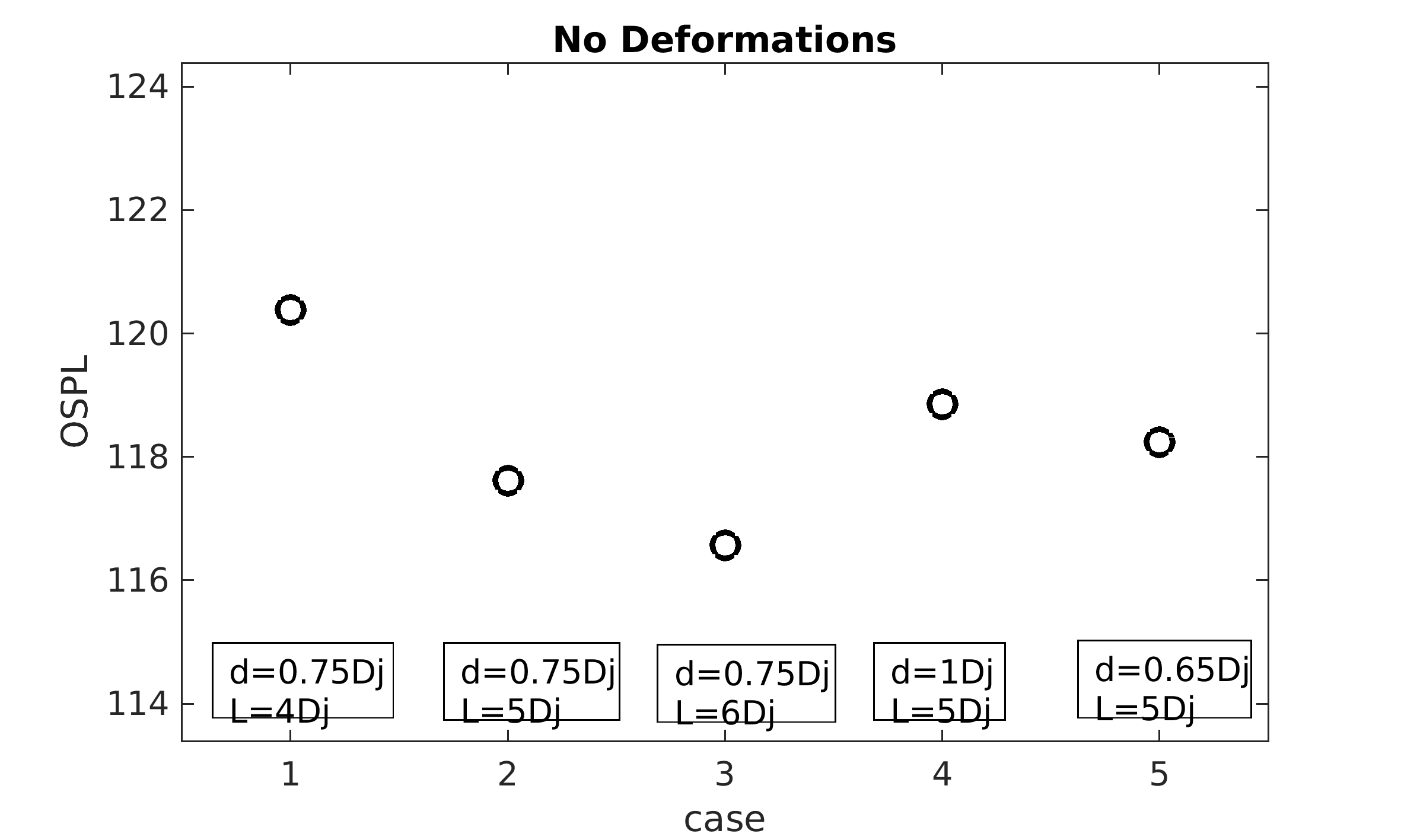}}}
	\end{tabular}
	\caption{Level of reduction of the overall sound pressure level for all probes compared to no edge deformations: (\textbf{a}) Probe 1; (\textbf{b}) Probe 2; (\textbf{c}) Probe 3.}
	\label{f16}
 \end{center}
\end{figure}

\begin{figure}[htp]
 \begin{center}
\begin{tabular}{cc}
	\subfigure[]{\includegraphics[width=0.45\textwidth]{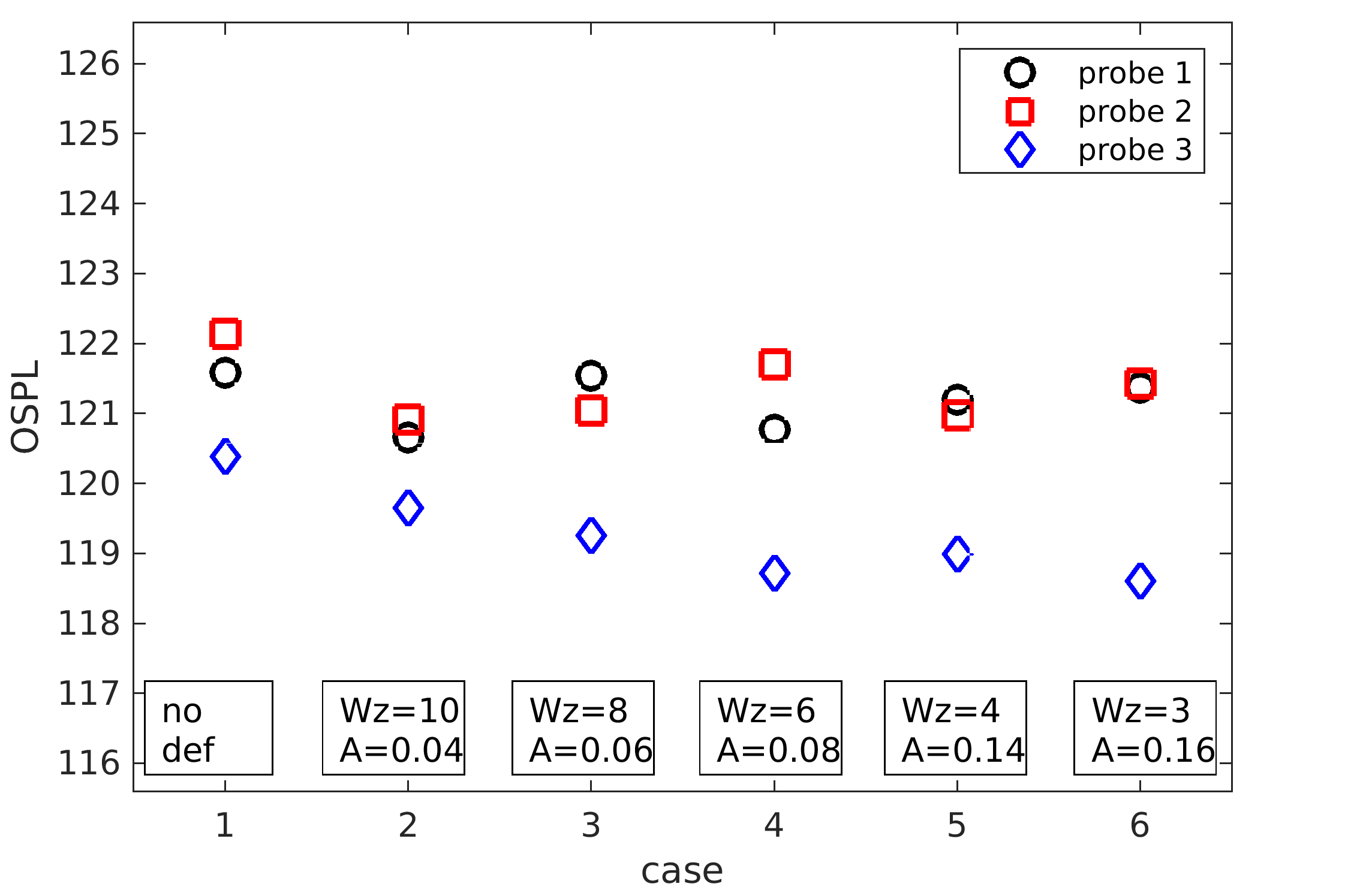}} &
	\subfigure[]{\includegraphics[width=0.45\textwidth]{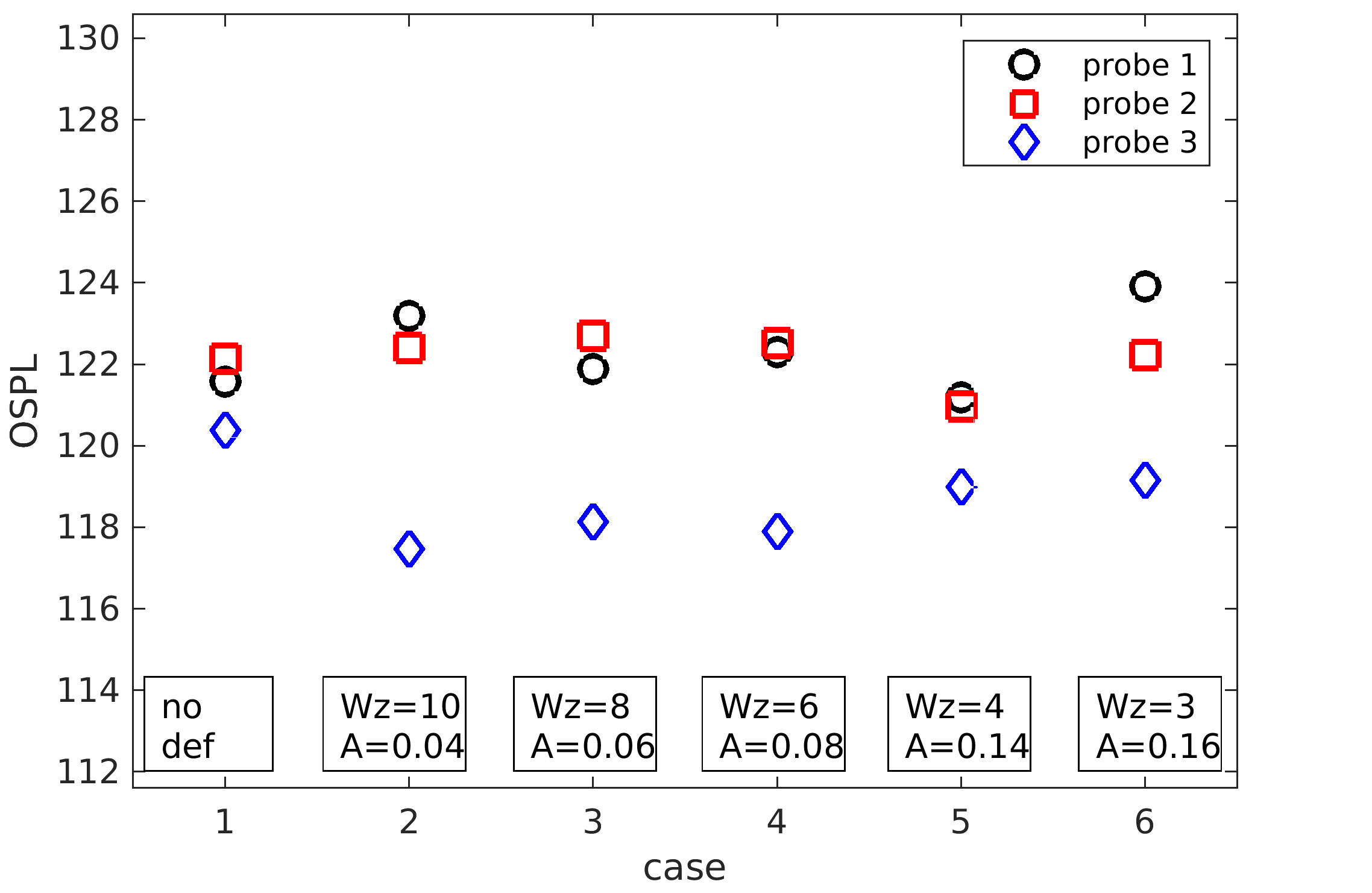}}\\
	\subfigure[]{\includegraphics[width=0.45\textwidth]{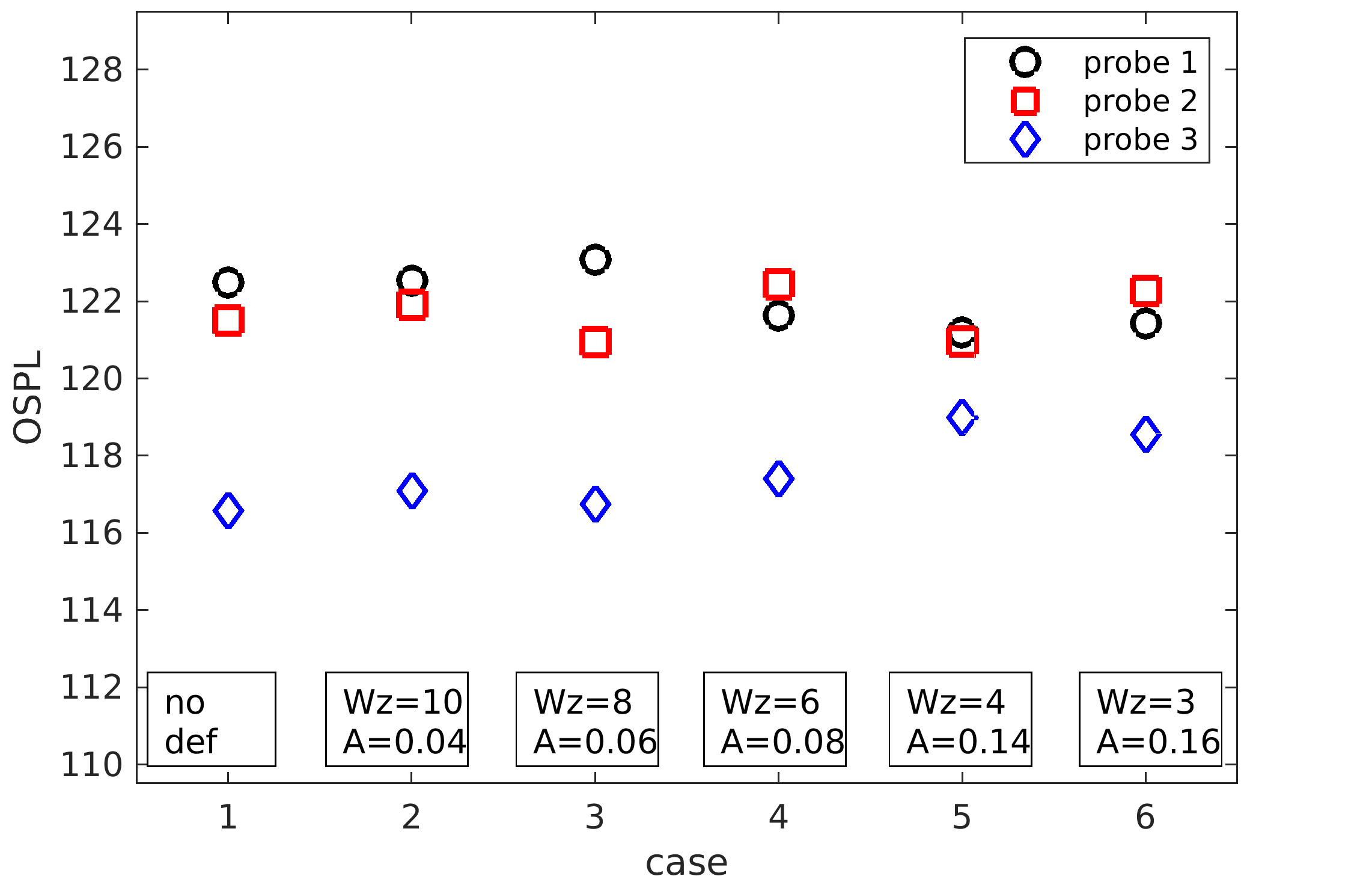} }&
	\subfigure[]{\includegraphics[width=0.45\textwidth]{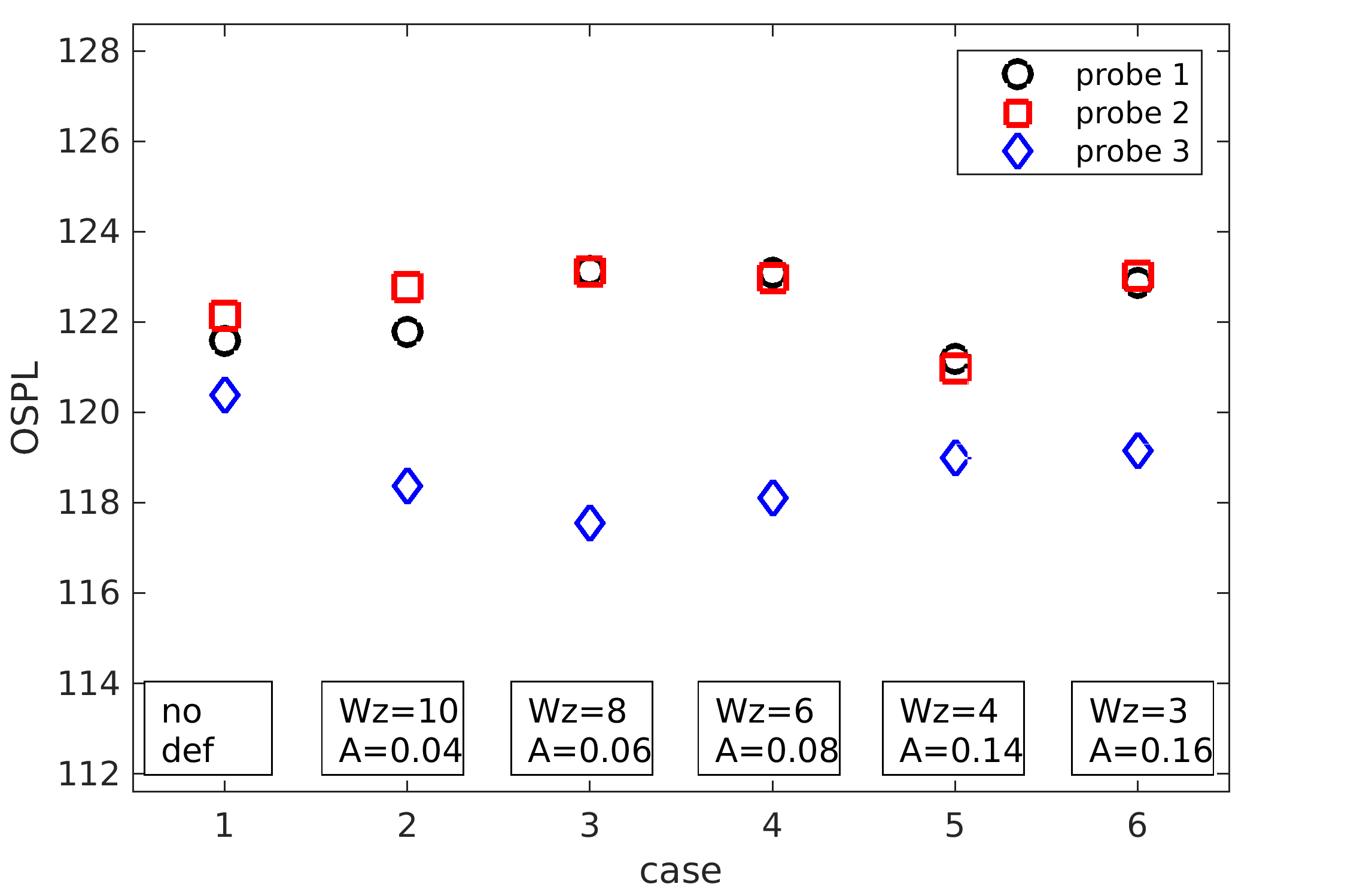}}\\
	\multicolumn{2}{c}{\subfigure[]{\includegraphics[width=0.45\textwidth]{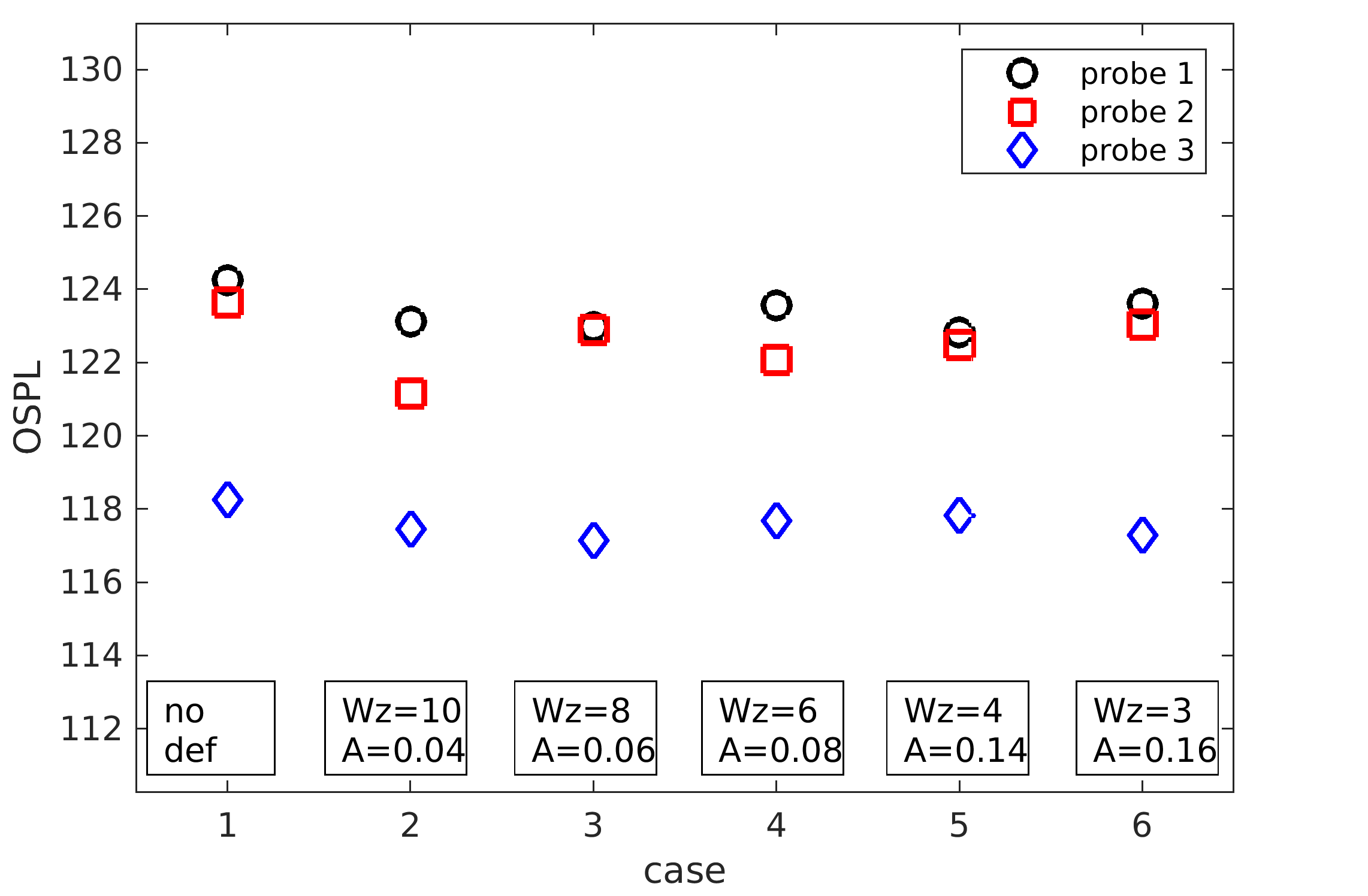} }}
	\end{tabular}
 \end{center}
	\caption{Overall sound pressure level for trailing edge deformations: (\textbf{a}) Case 1; (\textbf{b}) Case 2; (\textbf{c})~Case~3; (\textbf{d}) Case 4; (\textbf{e}) Case 5.}
	\label{f17}
\end{figure}

The overall sound pressure levels (OASPL) observed at the probe points are plotted in Figures~\ref{f16} and \ref{f17}. In Figure \ref{f16}, the OASPL's are plotted for the five configurations that do not involve trailing edge deformations. For the probes that are located above the jet, there is a significant increase in the radiated noise for the case corresponding to the smallest distance from the nozzle center ($d = 0.65 D_j$, $L = 5 D_j$). This is to be expected for this case since the end of the potential core is the closest to the trailing edge of the plate. For the probe located under the jet, there is an increase in the radiated noise for the case corresponding to the shortest plate length. This is an indication of a reduced shielding~effect.

Figure~\ref{f17} displays the OASPL results {at all three probe locations} for each of the tested configurations and prescribed trailing edge deformations. Each plot displays the results for all of the treatments given in Table~\ref{t2} at one of the fixed plate configurations given in Table~\ref{t1}. When the data are broken down by each plate placement location, there were two apparent plate configurations that yielded a noticeable reduction in sound levels at the probe locations. We will look further into these configurations to determine if the trailing edge deformations provide an additional positive effect.

There appeared to be an overall reduction in noise for most of the treatments in case~1 ($d=0.75 D_j$ $L=4 D_j$). The average overall sound pressure level for the no deformation case was 121.367 dB. For the $w_z=10$ $A=0.04$ treatment, the average was 120.467 dB, and, for the $w_z=6$ $A=0.08$ case, the average was 120.4 dB. The results for the Overall Sound Pressure Level data can be seen in Table~\ref{t4}. Although there was a reduction in noise for all of the deformations at this plate placement location, the two deformations mentioned above demonstrated the greatest reduction in noise.

 \begin{table}[htp]
 \caption{Overall Sound Pressure Level results for the probes corresponding to the plate placement of $d$ = 0.75 $D_j$ $L$ = 4 $D_j$.}
  \label{t4}

 \begin{center}
 \begin{tabular}{cccc}  \hline
      \textbf{Probe} &  \textbf{No Def.} &  \textbf{wz = 10 A = 0.04}   & \textbf{wz = 6 A = 0.08}  \\   \hline
       1 &  121.5 &  120.7 & 120.8 \\ 
       2 &  122.1 &  121.0 & 121.6 \\
       3 &  120.5 &  119.7 & 118.8   \\   \hline
  \end{tabular}
 \end{center}
   \end{table} 

For case 5 ($d=0.65 D_j$ $L=5 D_j$), there was also a reduction in noise for all of the deformations applied to the trailing edge of the plate. The average overall sound pressure level for the probes for the no deformation case was 122.067 dB. For the $w_z=10$ $A=0.04$ treatment, the average was 120.633 dB, and for the $w_z=4$ $A=0.14$ treatment the average was 121.133 dB. The Overall Sound Pressure Level results for this plate placement location can be seen in Table~\ref{t5}.

\begin{table}[htp]
 
   \caption{Overall Sound Pressure Level results for the probes corresponding to the plate placement of $d$ = 0.65 $D_j$ $L$ = 5 $D_j$.}
  \label{t5}
 \begin{center}
 \begin{tabular}{cccc}
   \hline
      \textbf{Probe} &  \textbf{No Def.} &  \textbf{wz = 10 A = 0.04}   & \textbf{wz = 4 A = 0.14}   \\  \hline
       1 &  124.2 &  123.2 & 122.9 \\ 
       2 &  123.7 &  121.3 & 122.6 \\ 
       3 &  118.3 &  117.4 & 117.9   \\   \hline
  \end{tabular}
 \end{center}
 \end{table}

The OASPL results from these configurations yielded the best results for overall noise reduction among those tested. It would be worthwhile to consider other deformations that are close to these treatments as well as other nearby plate placements to determine if further noise reduction could be achieved.



\section{Conclusions}

In this work, jet flow over various surface deformations at the trailing edge of a plate installed under a high aspect ratio rectangular jet were simulated to determine the effect of the jet-surface interaction on the noise radiated to the farfield. RANS calculations and a suite of LES simulations were performed with a high-order accurate flow solver. Numerous configurations and flow conditions were investigated, and the emitted noise from the jet-surface interactions calculated using Ffowcs-Williams and the Hawkings acoustic analogy method.

The RANS results, consisting of centerline velocity distributions, showed that the presence of the plate appeared to slow down the flow inside the potential core. For some of the tested cases, a slight acceleration of the flow further in the downstream was observed. The amount of acceleration appears to depend on the offset distance of the plate from the center of the jet, $d$. TKE distributions along the jet centerline revealed a decrease in TKE magnitude in the downstream region, which was expected. This decrease was more significant for the configuration with the smallest $d$ and smallest $L$. Cross-flow profiles of mean velocity magnitude and turbulent kinetic energy at different axial locations showed a clear deviation of the jet from the axis of symmetry near the trailing edge. The TKE profiles indicated that the jet is slightly skewed around the trailing edge of the plate in a manner resembling the Coanda effect. Despite this deviation, the cross-flow profile becomes more symmetric further downstream. A reduction of the TKE in the lower shear was observed for all cases with the plate under the jet. This is most significant for the configurations with the smallest distance to the trailing edge.

For the LES analysis, various results consisting of iso-surface of Q-criterion, centerline velocity, acoustic spectra, and overall sound pressure levels were presented and discussed. Overall, good agreement was found between the centerline velocities from RANS and LES, with some small discrepancies in the potential core region. Acoustic spectra plotted for the configurations without trailing edge deformations confirmed a reduction of the noise under the plate from the shielding effect. For most cases involving the trailing edge deformations, there did not appear to be a significant change to overall sound levels. However, some cases did demonstrate a noticeable increase or decrease in emitted noise. In the best cases, the overall sound pressure levels were decreased from 1 to 5 dB.

The results presented in this work provide evidence that further refinement of the configurations described herein may yield additional noise reductions. Future research aimed at analyzing the interactions of the jet flow with different types of surface deformations (e.g., non-periodic and/or streamwise-oriented) may uncover a greater degree of noise reduction, beyond what our results have shown.
\vspace{6pt}


%


\end{document}